\RequirePackage{lineno} 
\documentclass[a4paper,10pt]{article}

\usepackage{amsmath}
\usepackage{amsthm}
\usepackage{amsfonts}
\usepackage{hyperref}
\usepackage{color}
\usepackage{multirow}
\usepackage[dvips]{graphicx}
\usepackage{subfigure}
\usepackage{amssymb}
\usepackage{epsfig}

\usepackage{setspace}

\usepackage[numbers, sort&compress]{natbib}

\usepackage[normalem]{ulem}

\newcommand{\comment}[1]{}
\begin{document}

%\doublespacing

%\linenumbers

%\bibliographystyle{apsrev}

%\bibliographystyle{plain}

%\bibliographystyle{achemso}

%\bibliographystyle{plainnat}

\bibliographystyle{unsrt} 

%\title{\textbf{Phase field simulation of two-phase incompressible flows using general pressure equation}}
%\title{\textbf{General pressure equation based simulations of two-phase incompressible flows}}
\title{\textbf{Numerical simulation of two-phase incompressible viscous flows using general pressure equation}}

%and efficient

\author{$\textbf{Jun-Jie Huang}^{1, 2\footnote{Corresponding author. E-mail: jjhuang1980@gmail.com; jjhuang@cqu.edu.cn.}}$\\
	\\
	$ ^1$  Department of Engineering Mechanics, College of Aerospace Engineering, \\
	Chongqing University, Chongqing 400044, China \\
	$ ^2$ Chongqing Key Laboratory of Heterogeneous Material Mechanics, \\
	Chongqing University, Chongqing 400044, China}

\maketitle

\textbf{Abstract}

The general pressure equation (GPE) is a new method proposed recently by Toutant (J. Comput. Phys., 374:822-842 (2018)) for incompressible flow simulation.
It circumvents the Poisson equation for the pressure and performs better than the classical artificial compressibility method. Here it is generalized for two-phase incompressible viscous flows with variable density and viscosity. First, the pressure evolution equation is modified to account for the density variation.
Second, customized discretizations are proposed to deal with the viscous stress terms with variable viscosity.
Besides, additional terms related to the bulk viscosity are included to stabilize the simulation.
The interface evolution and surface tension effects are handled by a phase-field model coupled with the GPE-based flow equations. The pressure and momentum equations are discretized on a stagger grid using the second order centered scheme and marched in time using the third order total variation diminishing Runge-Kutta scheme.
Several unsteady two-phase problems in two dimensional, 
axisymmetric and three dimensional geometries
at intermediate density and viscosity ratios 
were simulated and the results agreed well with those obtained by other incompressible solvers and/or the lattice-Boltzmann method (LBM) simulations.
Similar to the LBM, the proposed GPE-based method is fully explicit and easy to be parallelized.
Although slower than the LBM, it requires much less memory than the LBM.
Thus, it can be a good alternative to simulate two-phase flows with limited memory resource.


\section{Introduction}\label{sec:intro}

%\textbf{Main Text}
To simulate incompressible viscous flows, it is often required to solve a Poisson equation to find the pressure.
This step consumes a large portion of the simulation time 
and is also a hurdle for parallel and adaptive computation.
Over the years, some methods have been proposed to overcome this problem, e.g.,
Chorin's artificial compressibility method (ACM)~\cite{jcp67-acm}, 
the lattice Boltzmann method (LBM)~\cite{Chen1998}, 
the method of entropically damped form of artificial compressibility (EDAC)~\cite{pre13-edac},
the general pressure equation (GPE) based method~\cite{jcp18-gpe},
and some others~\cite{prl05-krlns, jcp10-acm, jcp12-lwacm}.
%circumvent this issue. %solution of Poisson equation.
All of them share the same fundamental idea of \emph{artificial compressibility (AC)}.
That is, the flow is not required to be strictly incompressible (which dictates a divergence free velocity field);
instead, a small degree of compressibility is allowed and the weakly compressible flow can approximate
the incompressible flow well enough at sufficiently low Mach (Ma) numbers.
Among them, the GPE based method is relatively new and has shown certain advantages 
(e.g., better accuracy and convergence) for some benchmark problems~\cite{jcp18-gpe}.
Recently, its capability to simulate turbulent flows was demonstrated in~\cite{jcp2020-gpe-turb, tfc2020-gpe-turb}. 
As far as we know, it has been used only for single phase flows till now.

To simulate two-phase flows, it is necessary to identify the interfaces between two fluids and consider the effects of the surface tension. Different methods have been invented to achieve such purposes, e.g.,
the methods of the volume-of-fluid~\cite{jcp81-vof}, front-tracking~\cite{FrontTracking1992}, level-set~\cite{jcp94-level-set} and phase-field (diffuse-interface)~\cite{DIMReview98, jacqmin99jcp}.
Within the broader AC framework (including the LBM and EDAC), 
there have been previous endeavors to simulate two-phase flows.
Various LB models for them have been developed with great success~\cite{ColorLBM91, PotentialLBM93, swift95, lee05:hdr}(see~\cite{haibo-book-mflbm} for review). 
However, it is well known that LBM uses many distribution functions in addition to the common macroscopic variables (the pressure and velocity) and demands more memory, especially for 3D problems.
Recently, some work has been done for two-phase flow simulations based on the EDAC method (combined with the diffuse-interface model)~\cite{ftc2020edac-di}. 
It requires much less memory %than the LBM 
and need not solve the Poisson equation.
Here we extend the GPE-based method~\cite{jcp18-gpe} for two-phase problems by combining a phase-field model for interface dynamics~\cite{jacqmin99jcp}.
Although bearing certain similarities to~\cite{ftc2020edac-di}, 
our work differs from it in several aspects. %from~\cite{ftc2020edac-di}.
%One major difference is that 
First, instead of a collocated grid in~\cite{ftc2020edac-di}, we use a staggered grid (as in~\cite{jcp18-gpe}) because it helps to overcome the checkerboard instability.
Second, %Another difference is that 
we include additional terms related to the bulk viscosity in the momentum equations  to stabilize the simulation following the idea in the LBM~\cite{dellar01pre} and some other AC method~\cite{jcp10-acm}.
Besides, the present pressure evolution equation extended from the GPE~\cite{jcp18-gpe} for two-phase flows
is closer to that recovered at the macroscopic scale by the LBM~\cite{he99:mp, jcp10lbm-drop-impact}
and is slightly different from that in~\cite{ftc2020edac-di} (the pressure convection term is omitted under low Ma here). From another perspective, our work shows one possible way to directly solve the macroscopic two-phase incompressible viscous flow equations (approximated by the two-phase LBM) using relatively simple discretization schemes.
The key factors to reach this goal include (1) to use a staggered grid (2) to use robust time marching such as the third order total variation diminishing (TVD) Runge-Kutta (RK) scheme (3) to include suitable diffusion terms in the pressure equation (4) to include certain numerical dissipation due to the bulk viscosity (5) to apply proper averaging of the pressure for its gradient evaluation. They are described in detail below.
%the density in the bulk region 

This paper is organized as follows.
Section \ref{sec:gov-eqns} presents the complete governing equations, 
including the modified pressure equation for two-phase flows with variable density and viscosity
and the modified momentum equations with bulk viscosity terms.
Section \ref{sec:discretizations} gives the detailed spatial discretizations of different terms
and also the time marching schemes.
Section \ref{sec:results} shows the results for several two-phase flow problems
and compares them with other reference results.
The effects of bulk viscosity and pressure averaging will be demonstrated in this part as well.
Section \ref{sec:conclusion} concludes this paper with some discussions on future work.

\section{Governing equations}\label{sec:gov-eqns}

The governing equations for single phase incompressible flow include 
the continuity and momentum equations,
\begin{equation}\label{eq:incompressible-div-free}
\boldsymbol{\nabla}  \cdot \boldsymbol{u} = 0 ,
\end{equation}
\begin{equation}\label{eq:incompressible-momentum}
\partial_{t} \boldsymbol{u} + \boldsymbol{u} \cdot \boldsymbol{\nabla}  \boldsymbol{u} = - \frac{1}{\rho} \boldsymbol{\nabla}  p + \nu \nabla^{2}  \boldsymbol{u},
\end{equation}
where $\boldsymbol{u} $ is the velocity vector, $\rho$ is the (constant) density, 
%(for simplicity, $\rho = 1$ is assumed for single phase flows), 
$p$ is the pressure and $\nu$ is the kinematic viscosity.
In the GPE-based simulations, the momentum equation, eq. \ref{eq:incompressible-momentum}, 
remains unchanged whereas
a pressure evolution equation replaces the continuity equation~\cite{jcp18-gpe},
\begin{equation}\label{eq:gpe-pressure}
\partial_{t} p + \rho c_{s}^{2}  \boldsymbol{\nabla}  \cdot \boldsymbol{u} = \nu \nabla^{2}  p .
\end{equation}
Here $c_{s}$ is the isothermal sound speed (just like that in the LBM, 
$c_{s}= \delta_{x} / (\sqrt{3} \delta_{t})$ with
$\delta_{x}$ and $\delta_{t}$ being the mesh size and time step).
Note that the pressure equation in~\cite{jcp18-gpe} involves an additional ratio $\gamma/Pr$ with $\gamma$ and $Pr$ being the heat capacity ratio and the Prandtl number, but they are numerical parameters that can be usually taken as $1$. Besides, eq. \ref{eq:gpe-pressure} is in dimensional form with $c_{s}^{2}$ replacing $1/Ma^{2}$ in the dimensionless form of~\cite{jcp18-gpe}.
Like other ACMs, GPE-based simulations do not require the velocity divergence to be always zero. 
Instead, it is allowed to vary in a small range.

For two-phase flows with variable density and viscosity, we use an order parameter function $\phi (\boldsymbol{x} , t)$ to distinguish the two fluids with $\phi = 1$ in one fluid (called "liquid") and $\phi = -1$ in the other fluid (called "gas"), and $\phi$ varies between $-1$ and $1$ in a narrow interfacial region that separates the two fluids. The densities and dynamic viscosities of the liquid and gas are $\rho_{L}$, $\rho_{G}$,
and  $\eta_{L}$, $\eta_{G}$, respectively.
The fluid density and dynamic viscosity are functions of $\phi$,
\begin{equation}\label{eq:density-viscosity-op-two-phase}
\rho(\phi) = 0.5 [ \rho_{L} (1 + \phi) + \rho_{G} (1 - \phi) ], \quad
\eta(\phi) = 0.5 [\eta_{L} (1 + \phi) + \eta_{G} (1 - \phi) ],
\end{equation}
and the kinematic viscosity is $\nu(\phi) = \eta(\phi) / \rho(\phi)$.
%In the present work, t
The following pressure and momentum equations are adopted for two-phase flows,
\begin{equation}\label{eq:gpe-pressure-two-phase}
\partial_{t} p + \rho(\phi) c_{s}^{2}  \boldsymbol{\nabla}  \cdot \boldsymbol{u} =  \boldsymbol{\nabla}  \cdot (\nu (\phi) \boldsymbol{\nabla}  p) ,
\end{equation}
\begin{equation}\label{eq:gpe-momentum-two-phase}
\rho(\phi) (\partial_{t} \boldsymbol{u} + \boldsymbol{u} \cdot \boldsymbol{\nabla}  \boldsymbol{u} ) = -  \boldsymbol{\nabla}  p + \boldsymbol{\nabla}  \cdot \boldsymbol{\tau} + \boldsymbol{F}_{ST}+ \rho (\phi) \boldsymbol{g},
\end{equation}
where $\boldsymbol{g}$ is a body force like gravity (assumed to be constant here), 
$\boldsymbol{F}_{ST}$ is the force due to surface tension (its specific form is given later), 
and $\boldsymbol{\tau}$ is the viscous stress tensor which takes the form,
\begin{equation}\label{eq:vis-stress-two-phase}
%\boldsymbol{\tau} = \eta (\phi) [\boldsymbol{\nabla}  \boldsymbol{u} + (\boldsymbol{\nabla}  \boldsymbol{u} )^{T}] +\eta_{b} (\phi) (\boldsymbol{\nabla}  \cdot \boldsymbol{u}) \boldsymbol{I}  ,
\boldsymbol{\tau} = \eta (\phi) [\boldsymbol{\nabla}  \boldsymbol{u} + (\boldsymbol{\nabla}  \boldsymbol{u} )^{T}] +\bigg( \eta^{\prime} (\phi) - \frac{2}{3} \eta (\phi) \bigg) (\boldsymbol{\nabla}  \cdot \boldsymbol{u}) \boldsymbol{I}  ,
\end{equation}
where $\boldsymbol{I}$ is the identity tensor of order two and $\eta^{\prime} (\phi)$ is the second dynamic viscosity~\cite{dellar01pre}.
Compared with the truly incompressible momentum equations, 
in eq. \ref{eq:vis-stress-two-phase} 
$\boldsymbol{\tau}$ contains an additional term proportional to $\boldsymbol{\nabla}  \cdot \boldsymbol{u}$.
This type of term also appears in other AC based methods (like the LBM~\cite{dellar01pre} and the link-wise ACM~\cite{jcp12-lwacm}).
It helps to stabilize the weakly compressible simulation.
In previous studies~\cite{dellar01pre, jcp12-lwacm}, 
the coefficient before $\boldsymbol{\nabla}  \cdot \boldsymbol{u}$ 
%in eq. \ref{eq:vis-stress-two-phase} 
was considered as an \emph{adjustable numerical parameter}
(for convenience,
we set $\eta_{b} (\phi) = \eta^{\prime} (\phi) - \frac{2}{3} \eta (\phi)$ and call it "bulk viscosity"  following~\cite{cercignani-book}).
Here, it is assumed to be the same as the shear viscosity, i.e., $\eta_{b} (\phi) = \eta (\phi)$,
which was found to be acceptable
as it makes the simulations stable while giving reasonably accurate results.
Note that there are different definitions of bulk viscosity (see~\cite{dellar01pre} for more elaborated discussions).
For constant body forces, one has $\rho(\phi) \boldsymbol{g} = (\rho(\phi) - \rho_{ref}) \boldsymbol{g} +  \rho_{ref} \boldsymbol{g}$ where the constant $\rho_{ref} \boldsymbol{g}$ 
can be absorbed into the pressure
as $- \boldsymbol{\nabla} p + \rho_{ref} \boldsymbol{g} = - \boldsymbol{\nabla} (p - \rho_{ref} \boldsymbol{g} \cdot \boldsymbol{r}) = - \boldsymbol{\nabla} p^{\prime}$~\cite{cmfd99book}. Here 
$\rho_{ref}$ is a reference density (can be chosen as $\rho_{G}$ or $\rho_{L}$ depending on the problem), $\boldsymbol{r} = x \boldsymbol{e}_{x} + y \boldsymbol{e}_{y} + z \boldsymbol{e}_{z}$ is the position vector and $p^{\prime}$ is a redefined pressure having the same form of evolution equation as $p$.
Then the momentum equation can be rewritten as,
\begin{equation}\label{eq:gpe-momentum-two-phase-const-bf}
\rho(\phi) (\partial_{t} \boldsymbol{u} + \boldsymbol{u} \cdot \boldsymbol{\nabla}  \boldsymbol{u} ) = -  \boldsymbol{\nabla}  p^{\prime} + \boldsymbol{\nabla}  \cdot \boldsymbol{\tau} + \boldsymbol{F}_{ST}
+ (\rho (\phi) -  \rho_{ref}) \boldsymbol{g},
\end{equation}
For conciseness
we drop the prime in $p^{\prime}$ below.

As to the pressure evolution equation,
when compared to eq. \ref{eq:gpe-pressure}, 
the left-hand-side (LHS) of eq. \ref{eq:gpe-pressure-two-phase} 
is modified to account for the variable density and
its right-hand-side (RHS) is modified to account for the variable kinematic viscosity.
Eq. \ref{eq:gpe-pressure-two-phase}  is similar to the pressure equation recovered at the macroscopic scale in  two-phase LBM simulations~\cite{he99:mp, jcp10lbm-drop-impact} except that an diffusion term appears on the RHS. Such diffusion of the pressure is important in GPE-based simulations~\cite{jcp18-gpe} and may also exist in LBM simulations~\cite{jcp2020mame} (it may be usually considered as high order terms and neglected).
Eq. \ref{eq:gpe-pressure-two-phase}   also resembles the pressure evolution equation in the EDAC-based simulations recently proposed in~\cite{ftc2020edac-di} except that there is no convection term on the LHS
(an additional switch of pressure diffusion was included in~\cite{ftc2020edac-di}, but it was found to be not necessary for the present method).
As mentioned in~\cite{jcp10lbm-drop-impact}, the convection of pressure has little effect when the flow speed is low ($Ma \ll 1$).

The order parameter is governed by the Cahn-Hilliard equation(CHE)~\cite{cahnhilliard58, jacqmin99jcp},
\begin{equation}\label{eq:che}
\partial_{t} \phi + \boldsymbol{u} \cdot \boldsymbol{\nabla}  \phi = M \nabla^{2} \mu ,
\end{equation}
where $M$ is the mobility, and $\mu$ is the chemical potential given by,
\begin{equation}\label{eq:chem-pot}
\mu = 4 a \phi (\phi^{2} - 1) - \kappa \nabla^{2} \phi .
\end{equation}
Here $a$ and $\kappa$ are two constants related to 
a (reference) surface tension $\sigma_{ref}$ (set to $1$) and interface width $W$ as 
$a = (3 \sigma_{ref} ) / (4 W)$, $\kappa = 3 \sigma_{ref} W / 8$.
The surface tension force term in eq. \ref{eq:gpe-momentum-two-phase} can be written as
$\boldsymbol{F}_{ST} = - \sigma \phi \boldsymbol{\nabla} \mu$
with $\sigma$ being the physical surface tension.
To summarize, in the present GPE-based simulations the two-phase flows are governed by the coupled equations for the pressure (eq. \ref{eq:gpe-pressure-two-phase}), momentum (eq. \ref{eq:gpe-momentum-two-phase}) and order parameter (eq. \ref{eq:che})  supplemented with the relations to find the density, viscosity and chemical potential (eqs. \ref{eq:density-viscosity-op-two-phase} and \ref{eq:chem-pot}).

The above governing equations were written in general form and applicable for problems in two dimensions (2D) and  three dimensions (3D). They may be also written using the cylindrical coordinates $(r, \theta, z)$ and simplified into a pseudo 2D form when the flow is symmetric about the $z-$axis ($\partial_{\theta} = 0$). If we assume there is no flow in the azimuthal direction ($u_{\theta} = 0$) and the body force acts only along the axis ($\boldsymbol{g} = g_{z} \boldsymbol{e}_{z}$ with $\boldsymbol{e}_{z}$ being the unit vector in $z-$direction), the axisymmetric %governing
pressure, momentum and Cahn-Hilliard equations read,
\begin{equation}\label{eq:gpe-pressure-two-phase-axisym}
\partial_{t} p + \rho(\phi) c_{s}^{2}  (\boldsymbol{\nabla}  \cdot \boldsymbol{u}) =  
\partial_{r} (\nu(\phi) \partial_{r} p) +  \partial_{z} (\nu(\phi) \partial_{z} p) +  \frac{\nu(\phi)}{r} \partial_{r} p,
\end{equation}
\begin{equation}\label{eq:gpe-momentum-two-phase-axisym-z}
\begin{split}
&\rho(\phi) [\partial_{t} u_{z} + (u_{z} \partial_{z} u_{z} +  u_{r} \partial_{r} u_{z}) ] \\
&= -  \partial_{z}  p 
+\partial_{z} [ 2\eta (\phi) \partial_{z} u_{z} ] + \partial_{r} [ \eta (\phi) (\partial_{r} u_{z} + \partial_{z} u_{r} ) ] + \partial_{z} [ \eta_{b} (\phi) (\boldsymbol{\nabla}  \cdot \boldsymbol{u}) ]
+ \frac{\eta (\phi)}{r} \partial_{r} u_{z}
 - \sigma \phi \partial_{z} \mu + \rho (\phi) g_{z},
\end{split}
\end{equation}
\begin{equation}\label{eq:gpe-momentum-two-phase-axisym-r}
\begin{split}
&\rho(\phi) [\partial_{t} u_{r} + (u_{z} \partial_{z} u_{r} +  u_{r} \partial_{r} u_{r}) ]\\
&= -  \partial_{r}  p 
 + \partial_{z} [ \eta (\phi) (\partial_{r} u_{z} + \partial_{z} u_{r} ) ] 
+ \partial_{r} [ 2\eta (\phi) \partial_{r} u_{r} ]
+ \partial_{r} [ \eta_{b} (\phi) (\boldsymbol{\nabla}  \cdot \boldsymbol{u}) ]
+ \frac{\eta (\phi)}{r} \bigg(\partial_{r} u_{r} - \frac{u_{r}}{r}\bigg)
 - \sigma \phi \partial_{r} \mu .
\end{split}
\end{equation}
\begin{equation}\label{eq:che-axisym}
\partial_{t} \phi + u_{z} \partial_{z} \phi + u_{r} \partial_{r} \phi 
= M \bigg( \partial_{z} \partial_{z} \mu + \partial_{r} \partial_{r} \mu + \frac{1}{r} \partial_{r} \mu\bigg) ,
\end{equation}
with the velocity divergence $\boldsymbol{\nabla}  \cdot \boldsymbol{u}$ 
and chemical potential $\mu$ given by, 
\begin{equation}\label{eq:div-v-axisym}
\boldsymbol{\nabla}  \cdot \boldsymbol{u} = 
 (\partial_{r} u_{r} + \partial_{z} u_{z}) + \frac{u_{r}}{r} ,
 \end{equation}
\begin{equation}\label{eq:chem-pot-axisym}
\mu = 4 a \phi (\phi^{2} - 1) - \kappa \bigg( \partial_{z} \partial_{z} \phi + \partial_{r} \partial_{r} \phi + \frac{1}{r} \partial_{r} \phi\bigg)  .
\end{equation}

\section{Spatial and temporal discretizations}\label{sec:discretizations}

\begin{figure}[htp]
	\centering
		\includegraphics[trim= 1mm 1mm 1mm 1mm, clip, scale = 0.75, angle = 0]{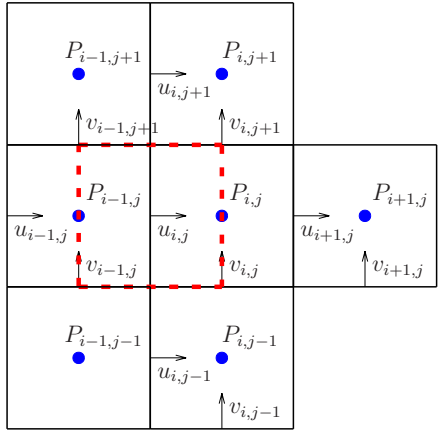}
	\caption{
		Staggered grid for spatial discretizations of the flow equations in 2D.
		The dashed lines form a typical cell for the discretization of the momentum equation (for $u$) along the horizontal direction.
	}
	\label{fig:staggered-grid}
\end{figure}

Following~\cite{jcp18-gpe}, we use a staggered uniform grid for the spatial discretizations of the flow equations due to its better performance to overcome the checkerboard instability.
Figure \ref{fig:staggered-grid} shows the arrangement of different variables in 2D).
The pressure is defined at the cell centers. 
The horizontal component of the velocity $u$ is defined at the middle points of the vertical edges
whereas the vertical velocity component $v$ is defined at the middle points of the horizontal edges.
The velocity divergence appears in both the pressure equation and the momentum equations
and it is evaluated at the cell centers. %(like the pressure).
Unless specified otherwise, the spatial derivatives are discretized by the second order centered scheme.
The formulas in 2D are given as an example (their extensions to 3D are straightforward). 
The velocity divergence is approximated by,
\begin{equation}\label{eq:div-v-discretization}
(\boldsymbol{\nabla}  \cdot \boldsymbol{u})_{i,j} \approx \frac{1}{\delta_{x}} [(u_{i+1,j} - u_{i,j}) + (v_{i,j+1} - v_{i,j}) ] .
\end{equation}
The diffusion term %with variable coefficient 
in eq. \ref{eq:gpe-pressure-two-phase}
may be written as $\boldsymbol{\nabla}  \cdot (\nu(\phi) \boldsymbol{\nabla} p) = \partial_{x} (\nu(\phi) \partial_{x} p) + \partial_{y} (\nu(\phi) \partial_{y} p)$
and the following scheme is used for $ \partial_{x} (\nu(\phi) \partial_{x} p) $,
\begin{equation}\label{eq:diff-p-discretization}
%\begin{split}
[\partial_{x} (\nu(\phi) \partial_{x} p)]_{i,j} 
\approx 
\frac{1}{\delta_{x}} [ (\nu(\phi) \partial_{x} p)_{i+\frac{1}{2},j} - (\nu(\phi) \partial_{x} p)_{i-\frac{1}{2},j} ] 
\approx \frac{1}{\delta_{x}^{2}} [ \nu_{i+\frac{1}{2},j} (p_{i+1,j} - p_{i,j})
-  \nu_{i-\frac{1}{2},j} (p_{i,j} - p_{i-1,j}) ] ,
%\end{split}
\end{equation}
where the kinematic viscosity at the cell edge is given by $\nu_{i\pm\frac{1}{2},j} = 0.5 [\nu(\phi_{i,j}) + \nu(\phi_{i\pm1,j})]$.
The momentum equations for $u$ and $v$ are discretized at different places due to the use of stagger grid.
The discretizations for $u$ are provided here (those for $v$ are similar).
The convection terms for $u$, $u \partial_{x} u + v \partial_{y} u$, are discretized as,
\begin{equation}\label{eq:conv-u-discretization}
\overline{(u \partial_{x} u + v \partial_{y} u)}_{i,j} \approx u_{i,j} \frac{u_{i+1,j} - u_{i-1,j}}{2 \delta_{x}} + \overline{v}_{i,j} \frac{u_{i,j+1} - u_{i,j-1}}{2 \delta_{x}} ,
\end{equation}
where $\overline{v}_{i,j} = (v_{i,j} + v_{i-1,j} + v_{i,j+1} + v_{i-1,j+1} ) / 4$.
Note that the indices for the overlined terms are the same as $u$
(i.e., at the vertical edge centers rather than at the cell centers).
In 2D, 
the $x-$component of $ \boldsymbol{\nabla}  \cdot \boldsymbol{\tau}$ is given by
$ \partial_{x} [ 2\eta (\phi) \partial_{x} u ] + \partial_{y} [ \eta (\phi) (\partial_{y} u + \partial_{x} v ) ] + 
\partial_{x} [ \eta (\phi) (\boldsymbol{\nabla}  \cdot \boldsymbol{u}) ]$
and the different terms are discretized as, 
\begin{subequations}\label{eq:u-stress-discretization}
	\begin{equation}
	%\begin{split}
\overline{	\partial_{x} [ \eta (\phi) \partial_{x} u ] }_{i,j} 
	\approx 
	\frac{1}{\delta_{x}} [ \overline{(\eta (\phi) \partial_{x} u)}_{i+\frac{1}{2},j} - \overline{(\eta (\phi) \partial_{x} u)}_{i-\frac{1}{2},j} ] 
	\approx \frac{1}{\delta_{x}^{2}} [ \overline{\eta}_{i+\frac{1}{2},j} (u_{i+1,j} - u_{i,j})
	-  \overline{\eta}_{i-\frac{1}{2},j} (u_{i,j} - u_{i-1,j}) ] ,
	%\end{split}
	\end{equation}
	\begin{equation}
	%\begin{split}
\overline{	\partial_{y} [ \eta (\phi) \partial_{y} u ]} _{i,j} 
	\approx 
	\frac{1}{\delta_{x}} [\overline{(\eta (\phi) \partial_{y} u)}_{i,j+\frac{1}{2}} - \overline{(\eta (\phi) \partial_{y} u)}_{i,j-\frac{1}{2}} ] 
	\approx \frac{1}{\delta_{x}^{2}} [ \overline{\eta}_{i,j+\frac{1}{2}} (u_{i,j+1} - u_{i,j})
	-  \overline{\eta}_{i,j-\frac{1}{2}} (u_{i,j} - u_{i,j-1}) ] ,
	%\end{split}
	\end{equation}
		\begin{equation}
%	\begin{split}
\overline{	\partial_{y} [ \eta (\phi) \partial_{x} v ] }_{i,j} 
	%&
	\approx 
	\frac{1}{\delta_{x}} [\overline{(\eta (\phi) \partial_{x} v)}_{i,j+\frac{1}{2}} - \overline{(\eta (\phi) \partial_{x} v)}_{i,j-\frac{1}{2}} ] %\\
%&	
\approx \frac{1}{\delta_{x}^{2}} [ \overline{\eta}_{i,j+\frac{1}{2}} (v_{i,j+1} - v_{i-1,j+1})
	-  \overline{\eta}_{i,j-\frac{1}{2}} (v_{i,j} - v_{i-1,j}) ] ,
%	\end{split}
	\end{equation}
		\begin{equation}
%	\begin{split}
	\overline{	\partial_{x} [ \eta(\phi) \boldsymbol{\nabla}  \cdot \boldsymbol{u} ] }_{i,j} 
	\approx 
	\frac{1}{\delta_{x}} [ \overline{(\eta(\phi) \boldsymbol{\nabla}  \cdot \boldsymbol{u})}_{i+\frac{1}{2},j} - \overline{(\eta(\phi) \boldsymbol{\nabla}  \cdot \boldsymbol{u})}_{i-\frac{1}{2},j} ] 
= \frac{1}{\delta_{x}} [ \overline{\eta}_{i+\frac{1}{2},j} (\boldsymbol{\nabla}  \cdot \boldsymbol{u})_{i+1,j} - \overline{\eta}_{i-\frac{1}{2},j} (\boldsymbol{\nabla}  \cdot \boldsymbol{u})_{i,j}],
%	\end{split}
	\end{equation}
\end{subequations}
where $\overline{\eta}_{i+\frac{1}{2},j} = \eta_{i,j}$, $\overline{\eta}_{i-\frac{1}{2},j} = \eta_{i-1,j}$,
$\overline{\eta}_{i,j \pm \frac{1}{2}} = (\eta_{i,j} + \eta_{i,j\pm1} + \eta_{i-1,j} + \eta_{i-1,j \pm 1}) / 4$.
The $x-$component of the surface tension force is discretized as,
\begin{equation}\label{eq:st-u-discretization}
\overline{(- \phi \partial_{x} \mu)}_{i,j} 
%\approx - \bigg[ \frac{1}{2 } (\phi_{i,j} + \phi_{i -1,j})  \bigg] \bigg[\frac{1}{\delta_{x}}(\mu_{i,j} - \mu_{i-1,j})  \bigg].
%\approx - [(\phi_{i,j} + \phi_{i -1,j}) / 2] [(\mu_{i,j} - \mu_{i-1,j}) / \delta_{x}] .
\approx - \frac{1}{2 \delta_{x}} (\phi_{i,j} + \phi_{i -1,j}) (\mu_{i,j} - \mu_{i-1,j})  .
\end{equation}
The pressure gradient term is discretized with some special treatment (similar, but not identical, to the isotropic scheme in~\cite{hybrid-mrt-lb-fd-axisym}),
\begin{equation}\label{eq:gradp-u-discretization}
\overline{( \partial_{x} p)}_{i,j} \approx \frac{1}{\delta_{x}}(\overline{p}_{i+\frac{1}{2},j} - \overline{p}_{i-\frac{1}{2},j} )  ,
\end{equation}
with $\overline{p}_{i + \frac{1}{2},j} = (4 p_{i, j} + p_{i, j+1} + p_{i, j-1})/6 $
and $\overline{p}_{i - \frac{1}{2},j} = (4 p_{i-1, j} + p_{i-1, j+1} + p_{i-1, j-1})/6 $.
That is, the pressure is first averaged before it is used to calculate the pressure gradient.
Note that in 3D the following formula is used to evaluate the averaged pressure 
(taking $\overline{p}_{i+\frac{1}{2},j, k}$ as an example),
\begin{equation}\label{eq:gradp-u-p-ave-3d}
\begin{split}
\overline{p}_{i+\frac{1}{2},j, k} = & [16 p_{i, j, k} + 4 (p_{i, j+1,k} + p_{i, j-1,k} + p_{i, j,k+1} + p_{i, j,k-1})\\
&+ (p_{i, j+1,k+1} + p_{i, j-1,k-1} + p_{i, j-1,k+1} + p_{i, j+1,k-1})] /36 .
\end{split}
\end{equation}
According to our experience, the above averaging of the pressure is important to stabilize 3D simulations 
%at high Reynolds numbers 
whereas it is not crucial for 2D cases.
With the above discretizations of the spatial derivatives, the pressure and velocity equations may be written in semi-discrete forms as,
\begin{subequations}\label{eq:p-u-v-semidiscrete-2d}
\begin{equation}
\frac{d p_{i,j}}{d t} = L_{p} (\rho_{i,j}, u_{i,j}, v_{i,j}, p_{i,j}) ,
\end{equation}
\begin{equation}
\frac{d u_{i,j}}{d t} = L_{u} (\rho_{i,j}, u_{i,j}, v_{i,j}, p_{i,j}, \phi_{i,j}, \mu_{i,j}) ,
\end{equation}
\begin{equation}
\frac{d v_{i,j}}{d t} = L_{v} (\rho_{i,j}, u_{i,j}, v_{i,j}, p_{i,j}, \phi_{i,j}, \mu_{i,j}) ,
\end{equation}
\end{subequations}
which are integrated in time by the third order TVD RK scheme (TVD-RK3)~\cite{jcp80-rk, moc98-tvd-rk}.
Suppose a quantity $q(t)$ follows the equation $dq / dt = L_{q} (q)$.
The steps to find $q^{n+1}$ at $t^{n} = (n+1) \delta_{t}$ from $q^{n}$ at $t^{n} = n \delta_{t}$ 
%$(p^{n+1}_{i,j}, u^{n+1}_{i,j}, v^{n+1}_{i,j})$ from $(p^{n}_{i,j}, u^{n}_{i,j}, v^{n}_{i,j})$
are as follows,
\begin{equation}\label{eq:tvd-rk3}
q^{(1)} = q^{n} + \delta_{t} L_{q} (q^{n}), \quad
q^{(2)} = \frac{3}{4} q^{n} + \frac{1}{4}  [ q^{(1)}  + \delta_{t}  L_{q} (q^{(1)})], \quad
q^{n+1} = \frac{1}{3} q^{n} + \frac{2}{3}  [ q^{(2)}  + \delta_{t}  L_{q} (q^{(2)})] .
\end{equation}
Note that $\phi$ and $\mu$ are defined at $t^{n+\frac{1}{2}} = (n + \frac{1}{2}) \delta_{t}$ and
during the integration of the flow equations they are assumed to be fixed (so are the density $\rho(\phi)$ and viscosity $\eta(\phi)$).

As to the CHE, the spatial derivatives in eq. \ref{eq:che} are discretized by 
the second order isotropic schemes
and its time marching uses the second order TVD RK scheme (TVD-RK2)~\cite{jcp80-rk, moc98-tvd-rk} %(for 2D and axisymmetric problems) 
or the fourth order RK scheme (RK4). %(for 3D problems).
%Take the 3D case as an example. 
The isotropic schemes to evaluate the spatial derivatives and Laplacian of $\phi$ read~\cite{hybrid-mrt-lb-fd-axisym},
\begin{equation}
\label{eq:1st-der-iso}
\partial_{\alpha} \phi
= \frac{3}{c \delta_{x}} \sum_{i=1}^{b} w_{i} e_{i \alpha} \phi (\boldsymbol{x} + \boldsymbol{e}_{i} \delta_{t})  ,
\end{equation}
\begin{equation}
\label{eq:laplacian-iso}
\nabla^{2} \phi
= \frac{6}{\delta_{x}^{2}} [\sum_{i=1}^{b} w_{i} \phi (\boldsymbol{x} + \boldsymbol{e}_{i} \delta_{t})
- (1 - w_{0}) \phi (\boldsymbol{x})]  .
\end{equation}
where $w_{i}$ is the weight associated with the lattice velocity $\boldsymbol{e}_{i}$, 
$\alpha = x$, $y$ or $z$, 
and $b$ is the total number of nonzero lattice velocity vectors (e.g., $b=18$ for the D3Q19 velocity model).
The details of $w_{i}$ and $\boldsymbol{e}_{i}$ for different velocity models may be found in the literature (e.g.,~\cite{jcp17-wmrt-lbm-mfflow}).
The steps to find $q^{n+1}$ from $q^{n}$ by the TVD-RK2 scheme are,
\begin{equation}\label{eq:tvd-rk2}
q^{(1)} = q^{n} + \delta_{t} L_{q} (q^{n}), \quad
q^{n+1} = \frac{1}{2} q^{n} + \frac{1}{2}   [ q^{(1)}  + \delta_{t}  L_{q} (q^{(1)})] ,
\end{equation}
and for the RK4 scheme,
\begin{subequations}\label{eq:rk4}
	\begin{equation*}
	a^{n} = \delta_{t} L_{q} (t^{n}, q^{n}), \quad
%	\end{equation*}
%\begin{equation*}
b^{n} = \delta_{t} L_{q} (t^{n} + \frac{1}{2} \delta_{t} , q^{n} + \frac{1}{2} a^{n}), 
\end{equation*}
\begin{equation*}
c^{n} = \delta_{t} L_{q} (t^{n} + \frac{1}{2} \delta_{t} , q^{n} + \frac{1}{2} b^{n}), \quad
%\end{equation*}
 %\begin{equation*}
 d^{n} = \delta_{t} L_{q} (t^{n} + \delta_{t} , q^{n} + c^{n}),
 \end{equation*}
\begin{equation*}
q^{n+1} =  q^{n} + \frac{1}{6}   (a^{n}  + 2 b^{n} + 2 c^{n} + d^{n}) .
\end{equation*}
\end{subequations}
To solve the CHE, the velocities at the cell centers are required and
they are obtained by simple averaging in both space and time. 
For example, in 2D the convection terms in the CHE can be expressed as 
$\boldsymbol{u} \cdot \boldsymbol{\nabla}  \phi = u \partial_{x} \phi + v \partial_{y} \phi$,
and the $x-$component velocity at the cell center (labelled with a hat) with indices $(i,j)$ at $t = (n+\frac{1}{2}) \delta_{t}$ is calculated as,
\begin{equation}\label{eq:u-center-2d}
\hat{u}^{n+\frac{1}{2}}_{i,j} = \frac{1}{2} (\hat{u}_{i,j}^{n} + \hat{u}_{i,j}^{n+1}) 
= \frac{1}{4} [(u_{i,j}^{n} + u_{i+1,j}^{n})  + (u_{i,j}^{n+1} + u_{i+1,j}^{n+1})] .
\end{equation}

\section{Results and Discussions}\label{sec:results}

In this section, we present the results of several two-phase incompressible viscous flow problems by using the new GPE-based method and make comparisons with those in the literature and 
by the LBM under the same simulation settings. %(i.e., same $\delta_{x}$, $\delta_{t}$). 
For the LBM simulations, the interface equation is actually solved in almost the same way as in the present method and the LBM is just used to deal with the flow equations.
The LBM formulation for two-phase flows follows~\cite{jcp10lbm-drop-impact} with some simplifications.
For 2D (including axisymmetric) and 3D problems, the LBM uses the D2Q9 and D3Q19 velocity models respectively, and the multiple relaxation time (MRT) collision model~\cite{pre2000-lbe-theory} or its recent variant (the weighted MRT model~\cite{jcp17-wmrt-lbm-mfflow}).  
It is noted that the LBM codes were already validated through several tests previously~\cite{hybrid-mrt-lb-fd-axisym, epje18-wbc-curved, apl19-ci-jump-drop-energy}.
Uniform mesh and time step are used in all problems. 
For each problem, a characteristic length $L_{c}$ and characteristic velocity $U_{c}$ are chosen.
The characteristic length $L_{c}$ is discretized into $N_{L}$ segments ($\delta_{x} = L_{c}/N_{L}$).
The characteristic time $T_{c} = L_{c}/U_{c}$ is divided into $N_{t}$ segments ($\delta_{t} = T_{c}/N_{t}$).
The sound speed is $c_{s} = c / \sqrt{3} = \delta_{x} / (\sqrt{3} \delta_{t}) = [N_{t}/ (\sqrt{3} N_{x}) ] U_{c}$.
For two-phase flows with nonvanishing surface tension ($\sigma > 0$), one can derive a velocity scale from the surface tension and liquid dynamic viscosity as $\sigma / \eta_{L}$.
In the following cases with $\sigma > 0$, 
we always use $U_{c} = \sigma / \eta_{L}$ and $T_{c} = L_{c}/U_{c}$ when we set up the simulation parameters.
At the same time, other characteristic velocity and time may be used to scale the relevant quantities (as described later).
When the densities and viscosities of the liquid and gas are given, one can calculate the density ratio as
$r_{\rho} = \rho_{L}/\rho_{G}$ and the dynamic viscosity ratio as $r_{\eta} = \eta_{L} / \eta_{G}$
(the kinematic viscosity ratio $r_{\nu} = \nu_{L} / \nu_{G} = r_{\eta} / r_{\rho}$).
Without loss of generality, we set $\rho_{L} = 1$ throughout this work.
In all cases, the initial velocities are set to $0$ and the initial pressure is set to a constant (e.g., $(\rho_{L} + \rho_{G}) c_{s}^{2}$) everywhere.

For two-phase flow simulations using the CHE, there are two additional parameters: (1) the Cahn number $Cn = W/L_{c}$ (i.e., the interface thickness measured by the characteristic length) and (2) the Peclet number $Pe = U_{c} L_{c}^{2} / (M \sigma_{ref})$ (reflecting the relative importance of convection over diffusion in the CHE).
The $Cn$ number should be small enough so that the results are sufficiently close to the sharp interface limit~\cite{jfm10-sil-che-cl}. In most of the following simulations, $Cn \leq 0.1$.
It is noted that there exist different definitions of $Cn$ in the literature.
If the definition in~\cite{kim05:cstff, jcp07-dim-ldr} is adopted, $Cn = 0.1$ here would become $0.1/(4\sqrt{2}) \approx 0.0177$.
The $Pe$ number must be also in a suitable range and it is given later for each problem.
Another important parameter is the interface thickness measured in grid size $W/\delta_{x}$.
It must be large enough to resolve the profile of $\phi$ in the interfacial region
(we used $W = 3 \delta_{x}$ or $4 \delta_{x}$).

\subsection{Capillary wave in 2D}\label{ssec:cap-wave}

\begin{figure}[htp]
	\centering
	\includegraphics[trim= 1mm 1mm 1mm 1mm, clip, scale = 0.65, angle = 0]{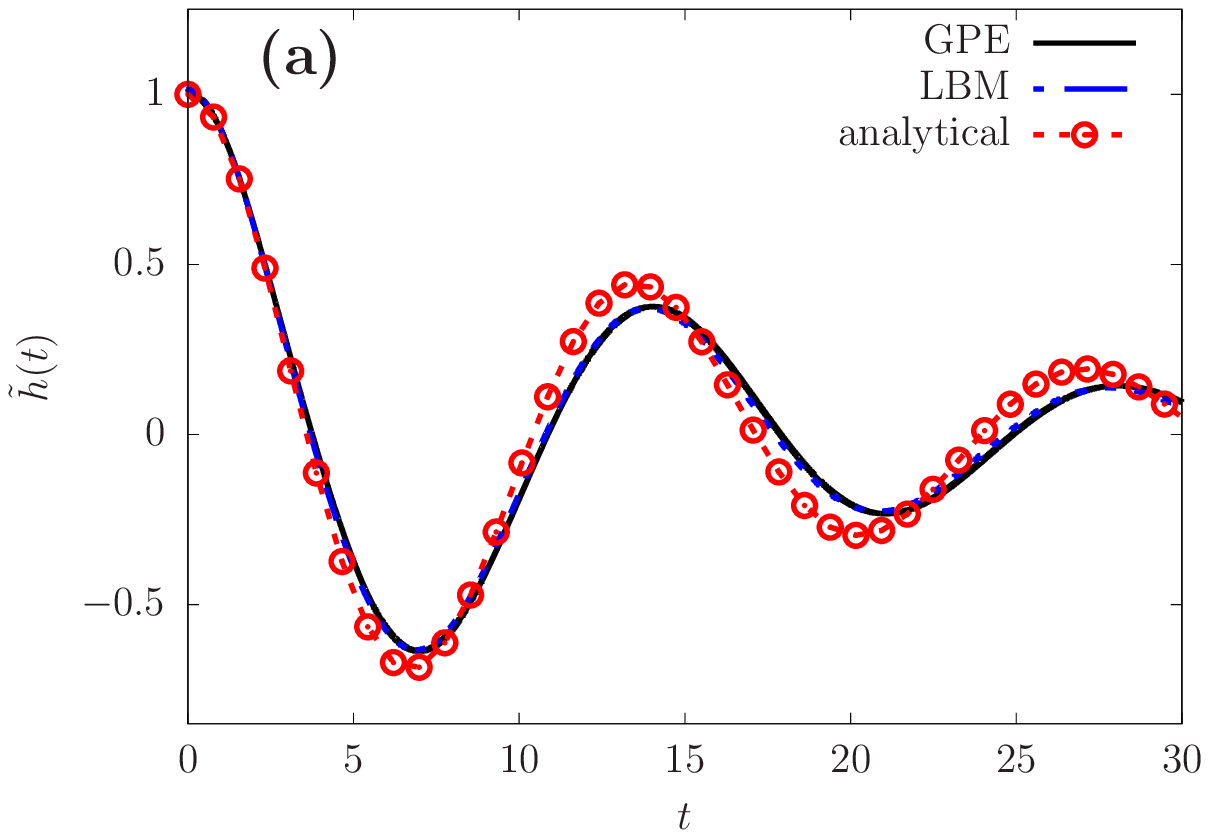}
	\includegraphics[trim= 1mm 1mm 1mm 1mm, clip, scale = 0.65, angle = 0]{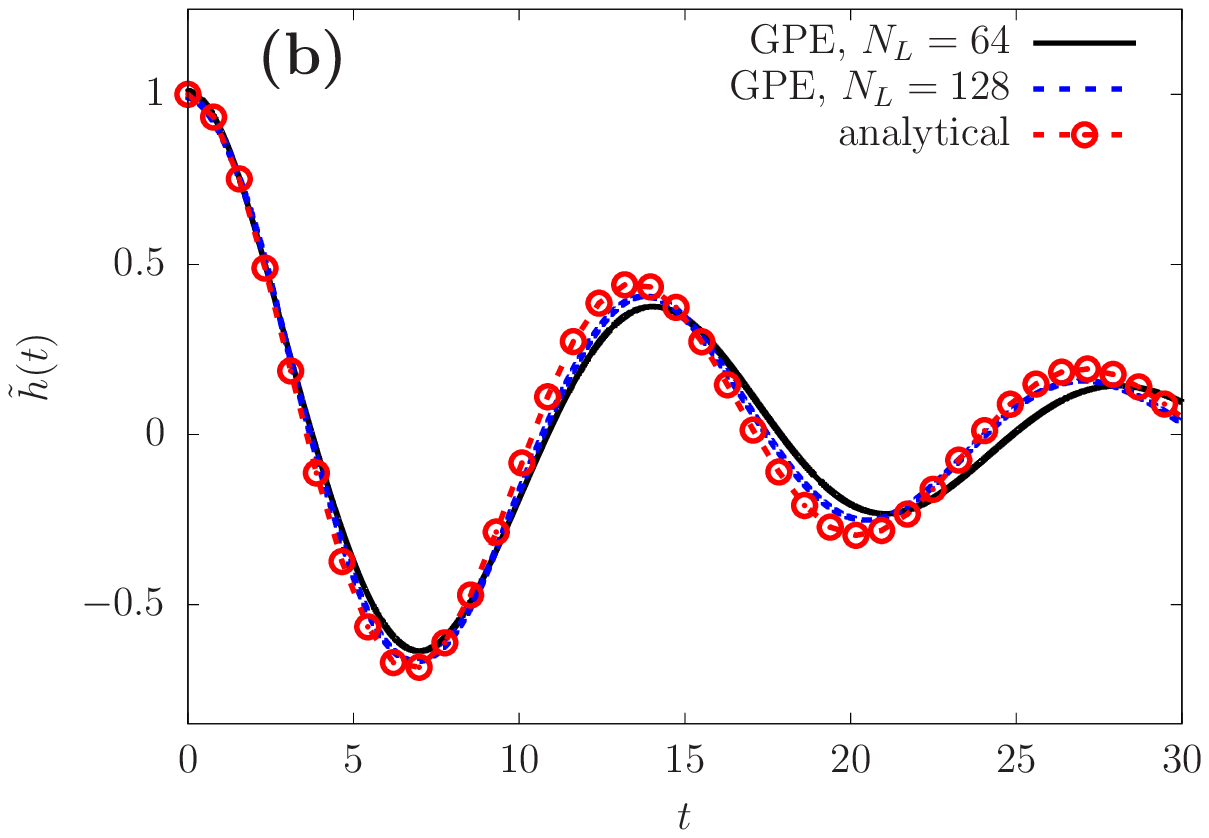}
	\caption{
		Evolutions of the (scaled) deviation of interface position from its equilibrium value measured at the lower boundary at $y = 0$ for the capillary wave problem: (a) comparison with LBM and the analytical solution; (b) effect of mesh refinement in GPE-based simulations.
	   In (a)
%	   the solid line is by the present method, the dashed line is by the LBM %using MRT
%		and the dash-dot line is the prediction by the analytical solution.
		the shared simulation parameters are
		$N_{L} = 64$, $N_{t} = 384$ ($c = 6$), $Cn = 0.0625$, $Pe = 2 \times 10^{4}$.
		In (b) the simulation parameters using the fine mesh are 
		$N_{L} = 128$, $N_{t} = 1920$ ($c = 15$), $Cn = 0.03125$, $Pe = 4 \times 10^{4}$
		whereas those for the coarse mesh are the same as in (a).
		%For the LBE, the model parameters for MRT follow those in fig. 1 of~\cite{pre2000-lbe-theory}.
	}
	\label{fig:cmp-cap-wave-evol}
\end{figure}

The first problem is the capillary wave in 2D.
The physical domain is a unit square $[0, 1] \times [0, 1]$ ($L_{c}$ is chosen as the side length, $L_{x} = L_{y} = 1$).
%The left and right boundaries are periodic, and the top and bottom boundaries are no-slip walls.
The left and right boundaries are no-slip walls, and the top and bottom boundaries are periodic.
The Reynolds number is defined as $Re = \rho_{L} U_{c} L_{c} / \eta_{L}$.
The gas and liquid phases occupy the left and right half domains, respectively.
The initial interface is slightly perturbed with the interface position varying with $y$ as
$h (y) = h_{eq} + A_{p} \cos[ k (y + 0.5)]$.
Here $h_{eq} = 0.5$ is the equilibrium interface position,
$A_{p} = 0.01$ is the amplitude of disturbance and 
$k = 2 \pi / \lambda$ is the wavenumber
($\lambda = 1$ is the wavelength).
The initial order parameter field is set to
$\phi(x,y, 0) = \tanh [ 2 (x - h(y) )  / Cn]$.
The interface position $h_{0}(t)$ at $y=0$ was monitored during the simulation.
Because of the symmetry about the line $y = 0.5$, only the lower half domain ($0 \leq y \leq 0.5$) 
was actually used with symmetric boundary conditions applied at $y = 0$ and $y = 0.5$.
We focus on the case at $Re = 1000$, $r_{\rho} = r_{\eta} = 20$ ($r_{\nu} = 1$). %was investigated.
For this problem, there is a fundamental frequency $\omega_{0} = \sqrt{\sigma k^{3} / (\rho_{L} + \rho_{G})}$.
When both the liquid and gas have the same the kinematic viscosity ($\nu_{L} = \nu_{G} = \nu$) and the perturbation is small ($A_{p} \ll 1$),
one can obtain the analytical solution for this problem~\cite{phf81:gravcapwav, kim05:cstff},
\begin{equation}\label{eq:cap-wave-analytical}
\tilde{h} (t) = \frac{h_{eq} - h_{0} (t)}{A_{p}} 
%= \frac{h(t) - h_{eq}}{h(0) - h_{eq}} 
%a_{\textrm{cw}}(t') 
= \frac{4 (1 - 4 \beta) \bar{\epsilon}^{2} }{8 (1- 4 \beta) \bar{\epsilon}^{2} + 1 } \text{erfc} (\sqrt{\bar{\epsilon} t'})
+ \sum_{i = 1}^{4} \frac{z_{i} \omega_{0}^{2}}{Z_{i} (z_{i}^{2} - \bar{\epsilon} \omega_{0}) } 
\exp \bigg[\frac{(z_{i}^{2} - \bar{\epsilon} \omega_{0}) t'} {\omega_{0}}\bigg]  \text{erfc}\bigg(z_{i} \sqrt{\frac{t'}{\omega_{0}}}\bigg) ,
\end{equation}
where $t' = \omega_{0} t $ and $\bar{\epsilon} = \nu k^{2} / \omega_{0}$
%\begin{equation}
%t' = \omega_{0} t, \quad \bar{\epsilon} = \frac{\nu k^{2}}{\omega_{0}}
%\end{equation}
are the scaled time and dimensionless viscosity, 
$\beta = \rho_{L} \rho_{G} / (\rho_{L} + \rho_{G})^{2}$,
$z_{i}$ are the four roots of the algebraic equation
$z^{4} - 4 \beta \sqrt{\bar{\epsilon} \omega_{0}} z^{3} + 2 (1 - 6 \beta) \bar{\epsilon} \omega_{0} z^{2}
+ 4 (1 - 3 \beta) (\bar{\epsilon} \omega_{0})^{\frac{3}{2}} z + (1 - 4 \beta) (\bar{\epsilon} \omega_{0})^{2} + \omega_{0}^{2} = 0$
%\begin{equation}
%z^{4} - 4 \beta \sqrt{\bar{\epsilon} \omega_{0}} z^{3} + 2 (1 - 6 \beta) \bar{\epsilon} \omega_{0} z^{2}
%+ 4 (1 - 3 \beta) (\bar{\epsilon} \omega_{0})^{\frac{3}{2}} z + (1 - 4 \beta) (\bar{\epsilon} \omega_{0})^{2} + \omega_{0}^{2} = 0 ,
%\end{equation}
and $Z_{1} = (z_{2} - z_{1}) (z_{3} - z_{1}) (z_{4} - z_{1})$,
$Z_{2} = (z_{3} - z_{2}) (z_{4} - z_{2})(z_{1} - z_{2}) $,
$Z_{3} = (z_{4} - z_{3}) (z_{1} - z_{3}) (z_{2} - z_{3})$,
$Z_{4} = (z_{1} - z_{4}) (z_{2} - z_{4})(z_{3} - z_{4}) $.
%\begin{subequations}
%	\begin{equation}
%	Z_{1} = (z_{2} - z_{1}) (z_{3} - z_{1}) (z_{4} - z_{1})
%	\end{equation}
%	\begin{equation}
%	Z_{2} = (z_{3} - z_{2}) (z_{4} - z_{2})(z_{1} - z_{2}) 
%	\end{equation}
%	\begin{equation}
%	Z_{3} = (z_{4} - z_{3}) (z_{1} - z_{3}) (z_{2} - z_{3})
%	\end{equation}
%	\begin{equation}
%	Z_{4} = (z_{1} - z_{4}) (z_{2} - z_{4})(z_{3} - z_{4}) 
%	\end{equation}
%\end{subequations}
Figure \ref{fig:cmp-cap-wave-evol}
shows the evolutions of $\tilde{h} (t) $ %$(h(t) - h_{eq})/(h(0) - h_{eq})$ 
over $0 \leq t \leq 30$ by using the present GPE-based method and the LBM, 
together with the analytical prediction given by eq. \ref{eq:cap-wave-analytical}.
It is seen that the present numerical results are very close to the LBM results (almost overlapping). 
Both numerical solutions agree with the analytical one in the early stage and the deviations grow slowly with time.
The deviations remain to be small after about two oscillation periods and can be reduced by refining the mesh and time step (see fig. \ref{fig:cmp-cap-wave-evol}b).

\subsection{Rayleigh-Taylor instability in 2D}\label{ssec:rt-instab}

The second problem is the Rayleigh-Taylor instability in 2D.
This initial condition and boundary conditions are similar to the capillary wave problem
except that the gas and liquid phases occupy the right and left half domains, respectively.
The surface tension $\sigma$ is set to zero, as in~\cite{he99:mp, jcp00-proj-fem, jcp07-dim-ldr}.
A gravity force with a magnitude $g$ is applied along the $x-$direction on the liquid %on the left 
pointing towards the gas %on the right 
($\rho_{ref}  = \rho_{G}$ in eq. \ref{eq:gpe-momentum-two-phase-const-bf}).
The physical domain is a rectangle $[0, 4] \times [0, 1]$ ($L_{c}$ is chosen as the domain height, $L_{y} = 1$, and the domain length $L_{x} = 4$).
Due to symmetry, only the lower half domain is used with symmetric boundaries at both the top and bottom sides. The left and right boundaries are no-slip walls.
The initial interface position varies with $y$ as $h (y) = h_{eq} + A_{p} \cos[ k (y + 0.5)]$
with $h_{eq} = 2$, $A_{p} = 0.1$ and $k = 2 \pi$ (the wavelength $\lambda = L_{y}= 1$).
The initial order parameter field is set to
$\phi(x,y, 0) = - \tanh [ 2 (x - h(y) )  / Cn]$.
As $\sigma = 0$, the characteristic velocity for this problem is derived from the wavelength $\lambda = L_{y}$ and $g$ as $U_{c} = \sqrt{\lambda g}$ and the characteristic time is $T_{c} = L_{c} / U_{c} = \sqrt{\lambda / g}$.
In this problem, an important dimensionless number is the Atwood number defined as $At = (\rho_{L} - \rho_{G})/(\rho_{L}+\rho_{G}) = (r_{\rho}  - 1) / (r_{\rho}  + 1) $. 
Here we focus on $At = 0.5$ (which gives the density ratio $r_{\rho} = 3$).
Two cases were studied.
In one case, the kinematic viscosities of the liquid and gas are the same ($r_{\nu} = 1$, giving $r_{\eta} %= r_{\rho} 
= 3$), as in~\cite{he99:mp}.
In the other case, the dynamic viscosities are the same ($r_{\eta} = 1$, giving $r_{\nu} = 1/3$), as in~\cite{jcp00-proj-fem, jcp07-dim-ldr}.
The Reynolds number is defined as $Re = \rho_{L} U_{c} L_{c} / \eta_{L} = \sqrt{g \lambda^{3}} / \nu_{L}$.
For the two cases with $r_{\eta} = 3$ and $r_{\eta} = 1$, the Reynolds number is set to $Re=256$ and $3000$, respectively.
For the latter case ($Re=3000$, $r_{\eta} = 1$) another characteristic time $T_{c}^{\prime} = T_{c} / \sqrt{At}$ was used in~\cite{jcp00-proj-fem, jcp07-dim-ldr}. For convenience, we denote the time scaled by $T_{c}^{\prime} $ as $t^{\prime}$ and it is related to $t$ as $t^{\prime} =t \sqrt{At}$.
Figure \ref{fig:cmp-cap-wave-evol}
shows the evolutions of the interface positions at $y=0$ (bubble front)
and at $y=0.5$ (spike front) together with the respective results from~\cite{he99:mp} 
and~\cite{jcp07-dim-ldr}.
The results of~\cite{he99:mp} were obtained by two-phase LBM simulations
and those of~\cite{jcp07-dim-ldr} were obtained by finite difference solutions of the coupled incompressible NSCH equations (which strictly enforces the incompressibility condition).
It is seen from fig. \ref{fig:cmp-cap-wave-evol} that the present interface positions are very close to the other two sets of results. 
Figure \ref{fig:rt-instab-snapshots} shows a few selected snapshots of the interfaces for the two cases.
The development of the complex interfacial pattern in fig. \ref{fig:rt-instab-snapshots} also resembles those in~\cite{he99:mp} (see fig. 6 therein for $Re=256$)
and~\cite{jcp07-dim-ldr} (see fig. 6 therein for $Re=3000$).

\begin{figure}[htp]
	\centering
	\includegraphics[trim= 1mm 1mm 1mm 1mm, clip, scale = 0.6, angle = 0]{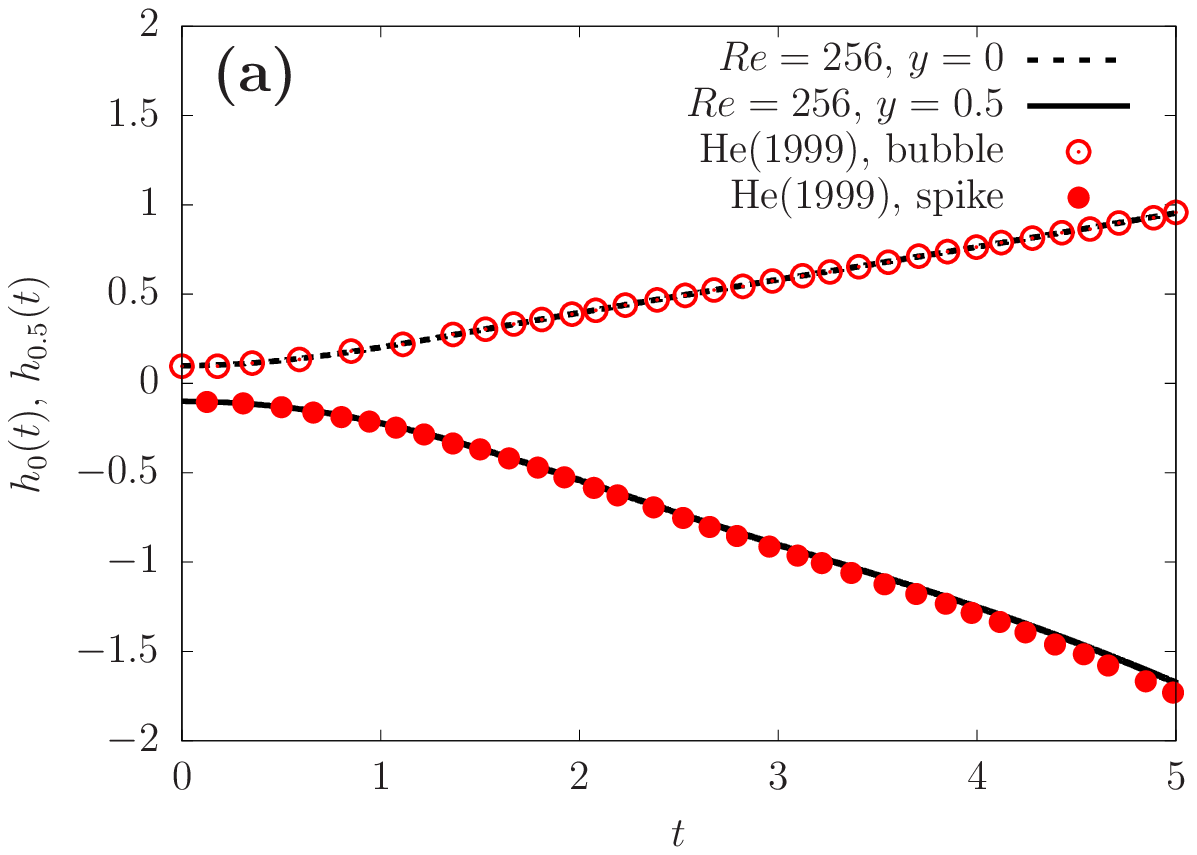}
\includegraphics[trim= 1mm 1mm 1mm 1mm, clip, scale = 0.6, angle = 0]{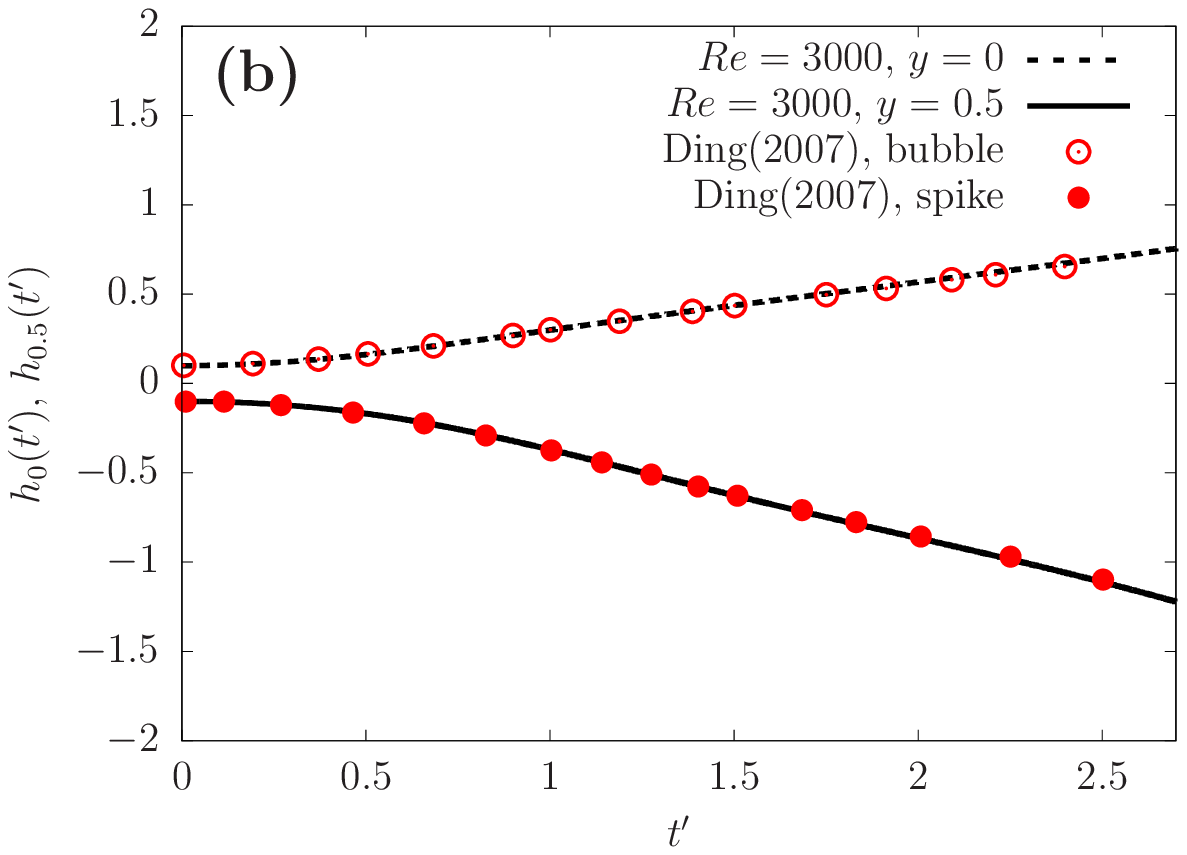}
	\caption{
		Evolutions of the interface positions at $y = 0$ (bubble front) and at $y=0.5$ (spike front) for the Rayleigh-Taylor instability problem
		at (a) $Re=256$, $r_{\eta} = 3$ %(case 1) 
		and (b) $Re=3000$, $r_{\eta} = 1$. %(case 2).
		The shared simulation parameters are
		$N_{L} = 256$, $N_{t} = 51200$ ($c = 200$),
		$Cn = 0.015625$, $Pe = 8 \times 10^{4}$.
	}
	\label{fig:rt-instab-itf-evol}
\end{figure}

\begin{figure}[htp]
	\centering
	\includegraphics[trim= 55mm 12mm 55mm 12mm, clip, scale = 1.2, angle = 0]{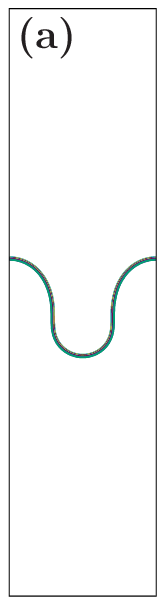}
	\includegraphics[trim= 55mm 12mm 55mm 12mm, clip, scale = 1.2, angle = 0]{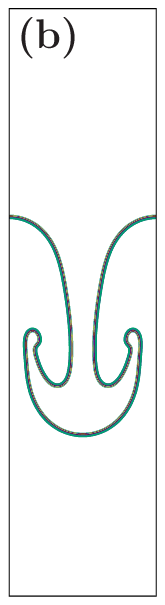}
	\includegraphics[trim= 55mm 12mm 55mm 12mm, clip, scale = 1.2, angle = 0]{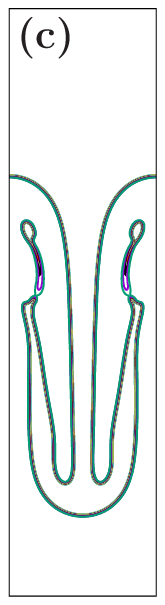}
		\includegraphics[trim= 55mm 12mm 55mm 12mm, clip, scale = 1.2, angle = 0]{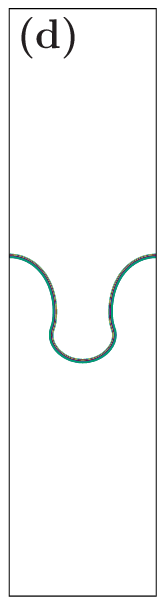}
			\includegraphics[trim= 55mm 12mm 55mm 12mm, clip, scale = 1.2, angle = 0]{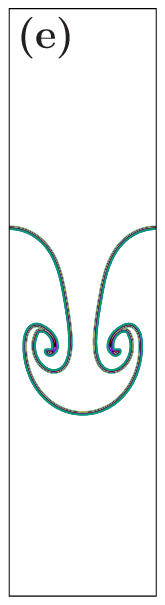}
				\includegraphics[trim= 55mm 12mm 55mm 12mm, clip, scale = 1.2, angle = 0]{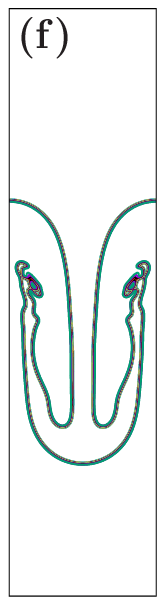}
	\caption{
		Snapshots of the interfaces at (a) $t = 1.5$ (b) $t=3.0$ and (c) $t =4.5$ for the first case at $Re = 256$ and $r_{\eta} = 3$ 
		and  (d) $t = 1.5$ ($t^{\prime} = 1.06$), (e) $t=2.5$ ($t^{\prime} = 1.77$) and (f) $t=3.5$ ($t^{\prime} = 2.47$) for the second case at $Re = 3000$ and $r_{\eta} = 1$ for the Rayleigh-Taylor instability problem by the present simulations.
		Note that the other half of the domain has been filled by using the symmetric conditions
		and the plots have been transformed to the usual setting with the gravity force pointing downwards.
	}
	\label{fig:rt-instab-snapshots}
\end{figure}

\subsection{Falling drop}\label{ssec:falling-drop}

\begin{figure}[htp]
	\centering
	\includegraphics[trim= 1mm 0mm 1mm 1mm, clip, scale = 0.65, angle = 0]{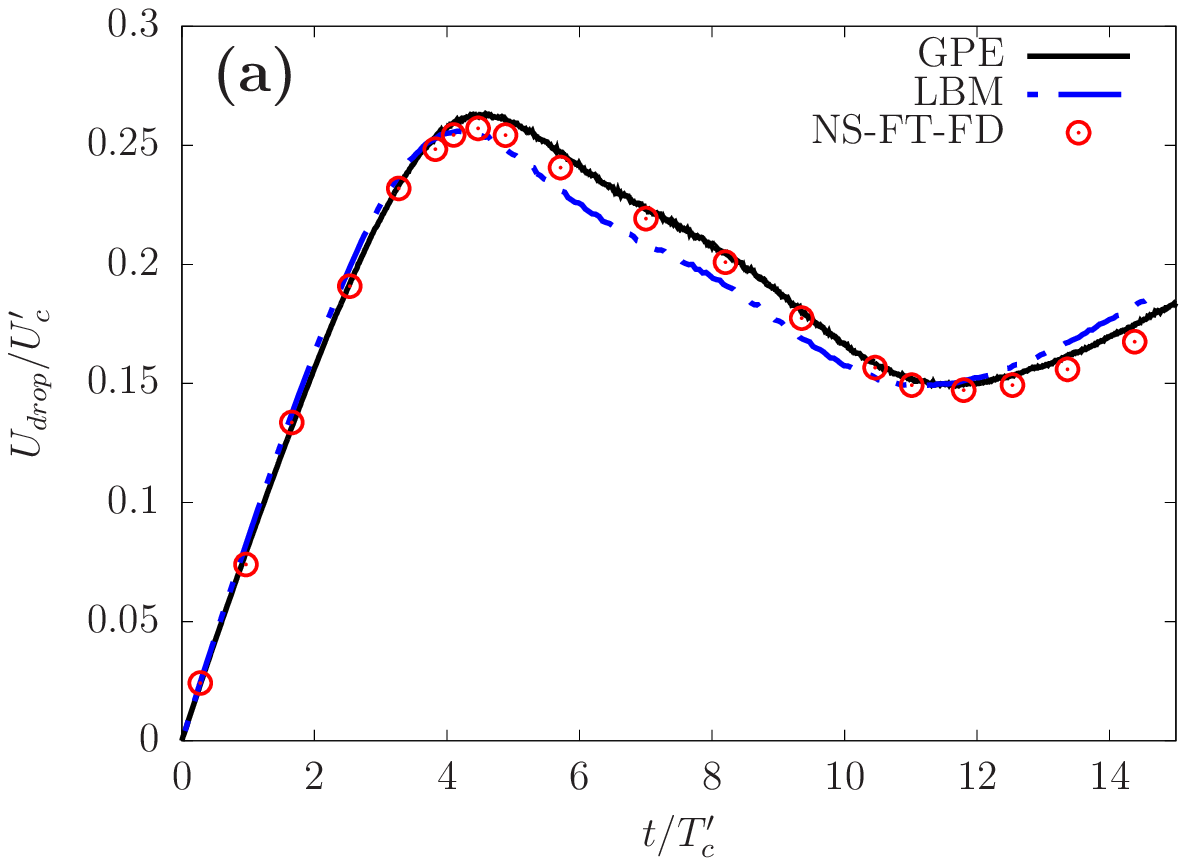}
	\includegraphics[trim= 1mm 0mm 1mm 1mm, clip, scale = 0.65, angle = 0]{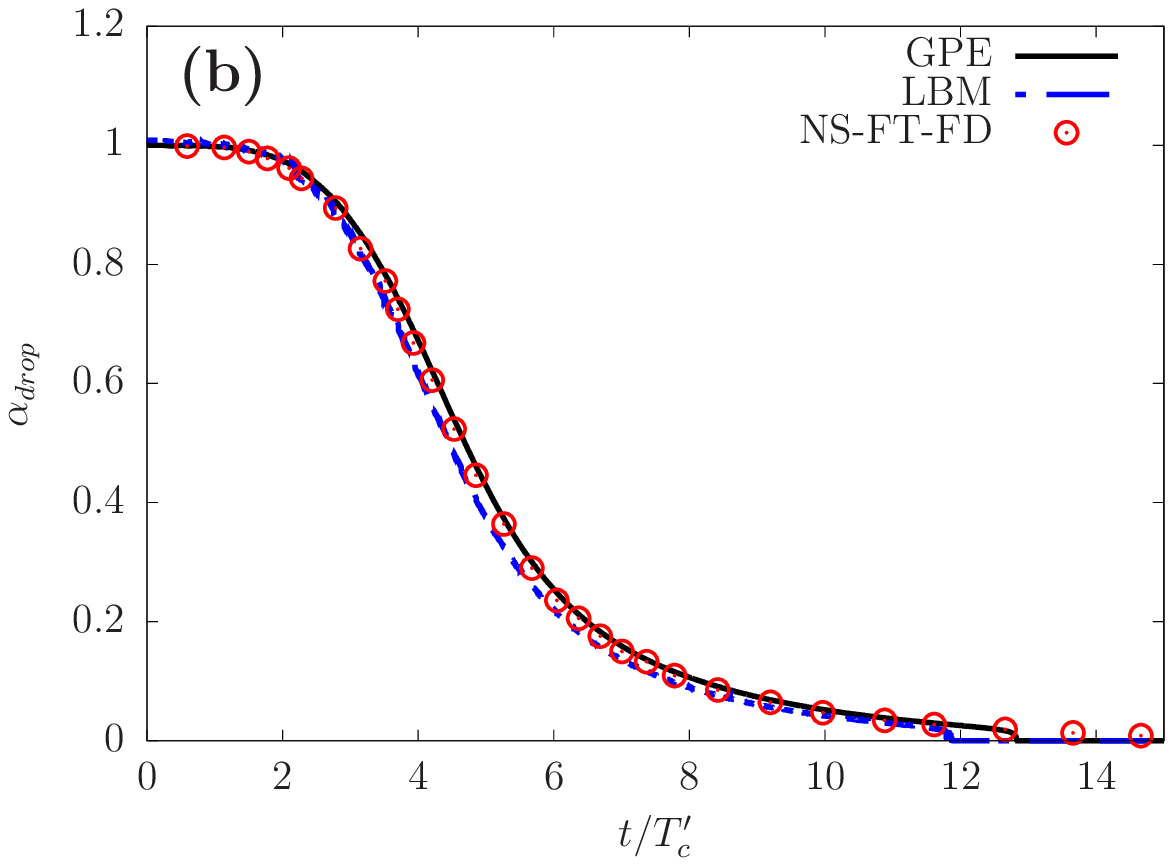}
	\caption{
		Evolutions of (a) the centroid velocity $U_{drop}$ in the $z-$direction and (b) the aspect ratio $\alpha_{drop}$ of the drop at  $Eo = 144$ and $Oh = 0.0466$ for the falling drop problem
		by the present GPE-based method, by the axisymmetric LBM in~\cite{hybrid-mrt-lb-fd-axisym} and by the finite difference solution of the NS equations and the front-tracking method~\cite{pof99dropacc}.
		The present simulation parameters are
		$N_{L} = 50$, $N_{t} = 4000$ ($c = 80$), $Cn=0.06$ and $Pe = 1000$.
	}
	\label{fig:falling-drop-evol}
\end{figure}

\begin{figure}[htp]
	\centering
	\includegraphics[trim= 35mm 17mm 16mm 6mm, clip, scale = 0.69, angle = 0]{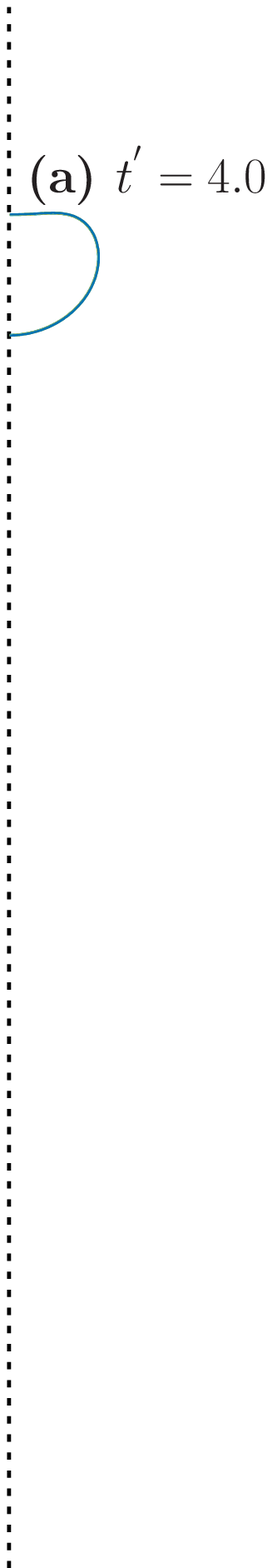}
	\includegraphics[trim= 35mm 17mm 16mm 6mm, clip, scale = 0.69, angle = 0]{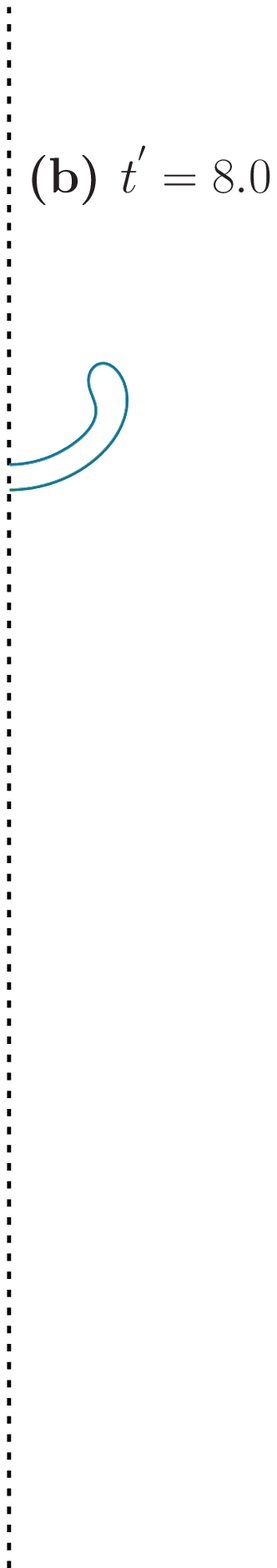}
    \includegraphics[trim= 35mm 17mm 16mm 6mm, clip, scale = 0.69, angle = 0]{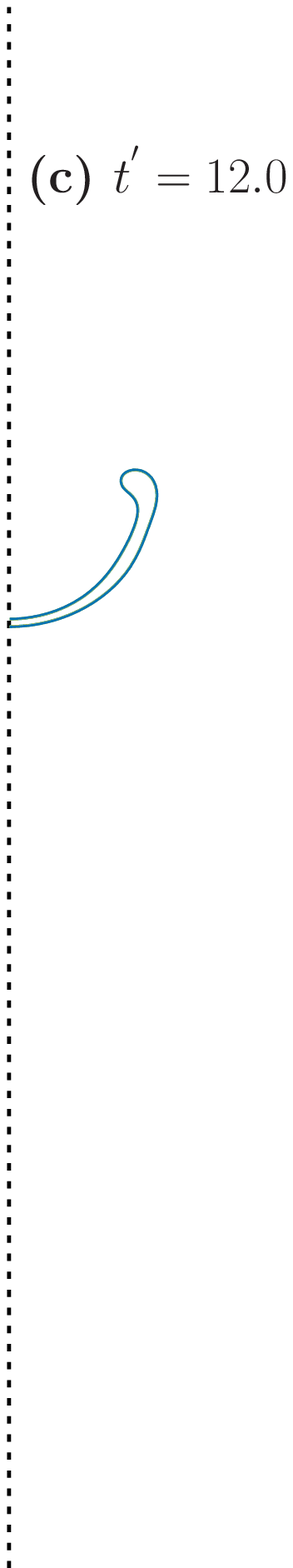}
	\caption{
		Snapshots of the interface at $t/T_{c}^{\prime} = 4.0$, $8.0$ and $12.0$ at $Eo = 144$ and $Oh = 0.0466$ for the falling drop problem by the present simulation.
	}
	\label{fig:falling-drop-itf-t04-08-12}
\end{figure}

The next problem is a falling drop under axisymmetric geometry.
It was studied by an axisymmetric LBM in~\cite{hybrid-mrt-lb-fd-axisym} and by a finite difference front tracing method in~\cite{pof99dropacc}.
In this problem, there is a drop surrounded by the ambient gas. 
The density ratio is $r_{\rho} = 1.15$
and the dynamic viscosity ratio is $r_{\eta} = 1$.
The drop radius $R$ is chosen as the characteristic length ($L_{c} = R$).
The domain is a rectangle $[0, 24] \times [0, 8]$ ($L_{z} = 24$ and $L_{r} = 8$).
Symmetric BCs were applied on the bottom side 
whereas no slip wall BCs were used for all other three boundaries.
Initially the drop center is at $(z_{c}, r_{c}) = (2, 0)$.
The initial order parameter field is set to
$\phi(z,r, 0) = - \tanh [ 2 (r_{dc} - R )  / Cn]$ where $r_{dc}= \sqrt{(z - z_{c})^{2} + (r - r_{c})^{2}}$.
The gravity force of magnitude $g$ is applied along the $z-$direction.
By choosing $\rho_{ref}  = \rho_{G}$ in eq. \ref{eq:gpe-momentum-two-phase-const-bf},
the gravity force was actually applied only on the drop.
Two main dimensionless parameters are the Eotvos number and Ohnesorge number defined as,
\begin{equation}\label{eq:falling-drop-dimless-num}
Eo = \frac{g (\rho_{L} - \rho_{G}) D^{2}}{\sigma}, \quad Oh = \frac{\eta_{L}}{\sqrt{\rho_{L} D \sigma}} ,
\end{equation}
where $D = 2R $ is the drop diameter.
They are set to $Eo = 144$ and $Oh = 0.0466$ (as in~\cite{hybrid-mrt-lb-fd-axisym, pof99dropacc}).
To facilitate the comparison with previous results, we scale the velocity and time using
$U_{c}^{\prime} = \sqrt{g D}$ and $T_{c}^{\prime} = \sqrt{D/g}$.
Note that the Boussinesq approximation was used in~\cite{hybrid-mrt-lb-fd-axisym}
under the relatively small density ratio. 
However, the present GPE-based method directly solves the momentum equations with variable density and viscosity, and there is no need to use such an approximation here.
The axisymmetric governing equations (eqs. \ref{eq:gpe-pressure-two-phase-axisym}, \ref{eq:gpe-momentum-two-phase-axisym-z}, \ref{eq:gpe-momentum-two-phase-axisym-r} and \ref{eq:che-axisym}) were solved. %with the following settings:
The centroid velocity $U_{drop}$ along the $z-$direction and 
the aspect ratio $\alpha_{drop} = Th_{drop} / Wh_{drop}$ of the drop 
were monitored during the simulation. Here $Th_{drop} $ and $ Wh_{drop}$ are the thickness (in the axial direction) and the width (in the radial direction) of the drop, respectively.
Note that $U_{drop}$ is calculated 
by $U_{drop} = \int_{A \vert_{\phi > 0}} r u(r,z)dr dz / \int_{A \vert_{\phi > 0}} r dr dz $ 
where $A \vert_{\phi > 0}$ represents the region where $\phi > 0$.
%\begin{equation}\label{falling-drop-U-drop}
%U_{drop} = \frac{\int_{A \vert_{\phi > 0}} r u(r,z)dr dz}{\int_{A \vert_{\phi > 0}} r dr dz} 
%\end{equation}
Figure \ref{fig:falling-drop-evol}
shows the evolutions of the centroid velocity in the axial direction and the aspect ratio of the drop obtained by the present method, together with those from~\cite{hybrid-mrt-lb-fd-axisym, pof99dropacc}.
It is seen that the three sets of results agree very well with each other in the early stage.
After some time, there are some small differences between them.
Overall, the present results are closer to that in~\cite{pof99dropacc} than the LBM results in~\cite{hybrid-mrt-lb-fd-axisym}
(probably because the present GPE-based simulation does not employ the Boussinesq approximation).
Figure \ref{fig:falling-drop-itf-t04-08-12} gives several snapshots of the interface.
They resemble the interfaces obtained by the other two methods %in~\cite{hybrid-mrt-lb-fd-axisym, pof99dropacc} 
as well (see fig. 2 in~\cite{pof99dropacc} and fig. 11 in~\cite{hybrid-mrt-lb-fd-axisym};
note the present selected times are close to those in the references, but not exactly the same).
%where $t^{\prime} = 3.87$, $7.73$ and $11.6$).

\subsection{Drop coalescence (axisymmetric)}\label{ssec:drop-coal}

\begin{figure}[htp]
	\centering
	\includegraphics[trim= 1mm 1mm 1mm 1mm, clip, scale = 0.65, angle = 0]{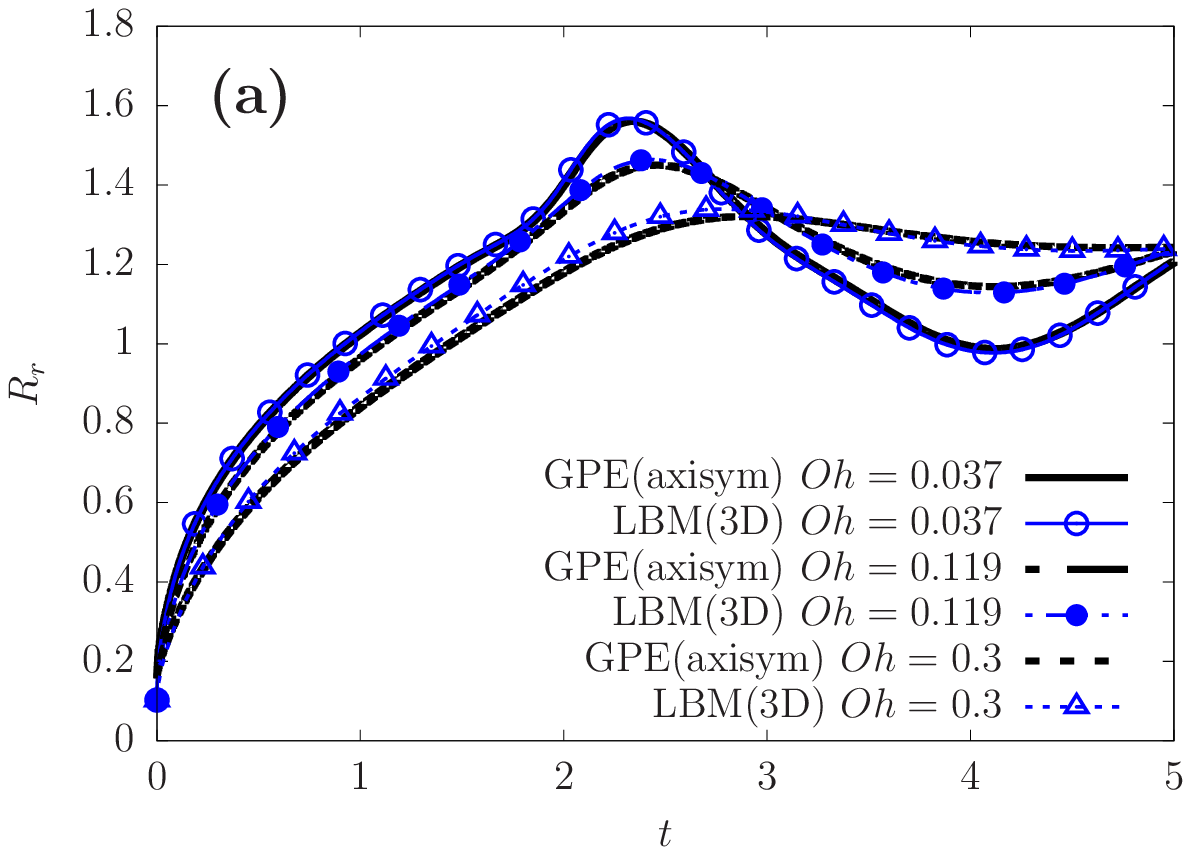}
	\includegraphics[trim= 1mm 1mm 1mm 1mm, clip, scale = 0.65, angle = 0]{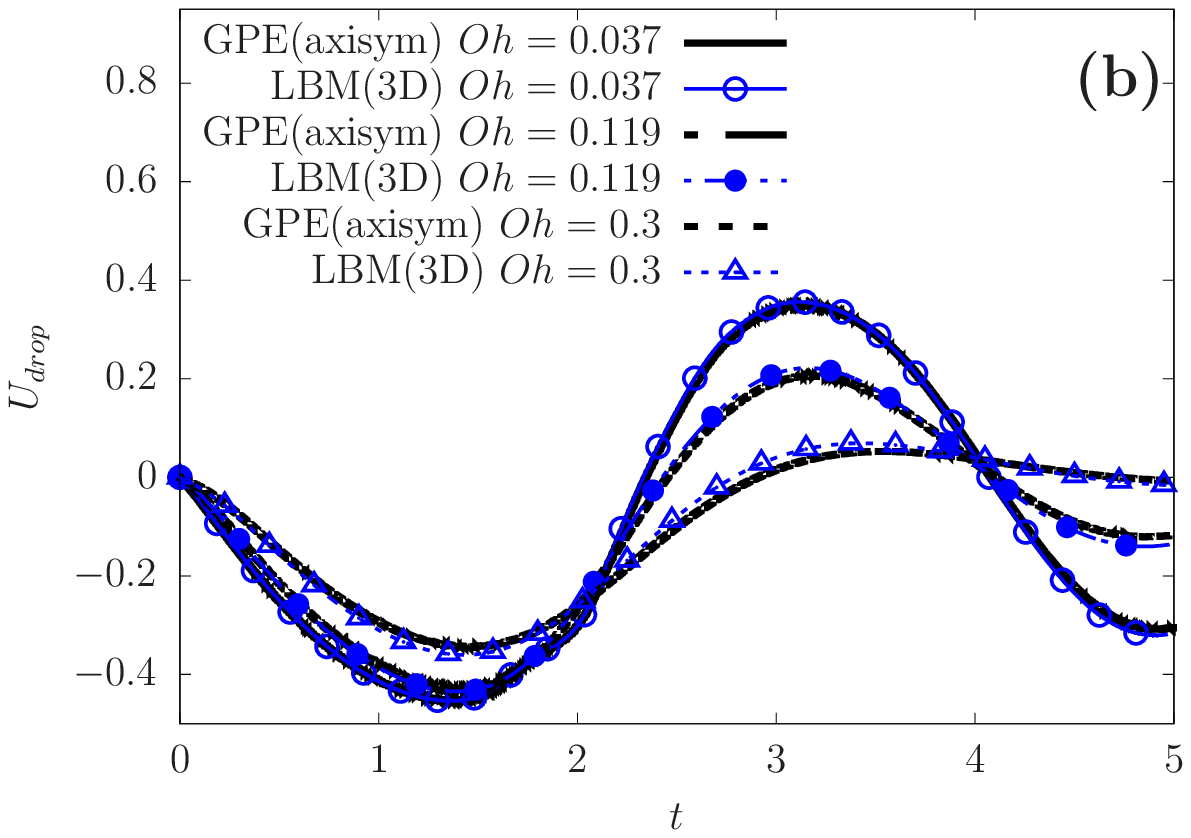}
	\caption{
		Evolutions of the radius of the drop in the radial direction at $z=0$ 
		and the axial component of the centroid velocity of the (right half) drop  
		during the coalescence of two drops by the present simulation.
	%	The upper row is for  $Oh = 0.3$, the middle for $Oh = 0.119$, and the lower, $Oh = 0.037$ .
		The shared simulation parameters are
		$N_{L} = 40$, 
		$Cn = 0.1$, $Pe = 8 \times 10^{3}$.
		The temporal discretization parameter is
		$N_{t} = 400$ ($c = 10$) for $Oh=0.119$ and $0.037$,
		and $N_{t} = 2000$ ($c = 50$) for $Oh=0.3$.
	}
	\label{fig:drop-coal-ry-udrop-evol}
\end{figure}

Next, we investigate the coalescence of two spherical drops of the same radius $R$ (chosen as the characteristic length $L_{c}$).
For this problem, one can define the capillary-inertial velocity
and time as $U_{ci} = \sqrt{\sigma / (\rho_{L} R)}$ and $T_{ci} = R/U_{ci} = \sqrt{\rho_{L} R^{3} / \sigma}$.
They are used to scale the relevant quantities.
The Ohnesorge number is defined as $Oh = \eta_{L} / \sqrt{\rho_{L} \sigma R} $.
%\begin{equation}\label{eq:oh-drop-coal}
%Oh = \frac{\eta_{L}}{\sqrt{\rho_{L} R \sigma}} 
%\end{equation}
Due to the symmetry about the axis connecting the two drop centers, 
the problem can be simplified as a 2D axisymmetric problem and only the upper and right quarter of the domain is used (symmetric boundary conditions are applied on all sides).
The domain is a  square $[0, 3] \times [0, 3]$ ($L_{z} =L_{r} = 3$).
The initial drop center is at $(z_{c}, r_{c} )= (1,0)$.
The initial $\phi$ field is set in the same way as in Section \ref{ssec:falling-drop}.
The density ratio is $r_{\rho} = 50$ and the dynamic viscosity ratio is $r_{\eta} = 58.8$ (same as in~\cite{jfm14jumping-drop-sim, apl19-ci-jump-drop-energy}).
Three cases at $Oh = 0.037$, $0.119$ and $0.3$ were simulated using the present GPE-based method (axisymmetric formulation).
During the simulations, the radius of the drop $R_{r}$ in the radial direction at $z=0$
and the centroid velocity of the (right half) drop in the $z-$direction $U_{drop}$
were monitored.
Figure \ref{fig:drop-coal-ry-udrop-evol} shows the evolutions 
of $R_{r}$ and $U_{drop}$ during the coalescence process.
The predictions by using the 3D LBM %using the D3Q19 velocity model 
as in~\cite{apl19-ci-jump-drop-energy} are also given for comparison.
It is seen that overall the present axisymmetric GPE simulations give predictions of the two quantities close to the 3D LBM results.

\subsection{Drop coalescence in 3D}\label{ssec:drop-coal-3d}

Lastly, a 3D problem is simulated by the proposed method. 
The problem is the same as that in Section \ref{ssec:drop-coal} except that it is handled under the original 3D setting without using the axisymmetric conditions.
The basic physical parameters are the same ($r_{\rho} = 50$, $r_{\eta} = 58.8$, $Oh = 0.037$, $0.119$ and $0.3$).
We still make simplifications by using some of the symmetric conditions.
The simulation domain ($[0, 3] \times [0, 3]\times [0, 3]$) corresponds to one eighth of the complete domain.
Symmetric BCs were applied on all the six boundary planes at $x = 0$, $3$, $y = 0$, $3$, and $z = 0$, $3$.
Initially the drop center is at $(x_{c}, y_{c}, z_{c}) = (1, 0, 0)$.
The initial order parameter field is set to
$\phi(x,y,z, 0) = - \tanh [ 2 (r_{dc} - R )  / Cn]$ where $r_{dc}= \sqrt{(x - x_{c})^{2} + (y - y_{c})^{2} + (z - z_{c})^{2}}$.
%Unlike Section \ref{ssec:drop-coal},
Here the half length of the drop on the $x-$axis $R_{x}$
and the centroid velocity of the drop (within the simulation box) in the $z-$direction $W_{drop}$
were monitored.
By default, the D3Q15 model was used in the isotropic schemes (eqs. \ref{eq:1st-der-iso} and \ref{eq:laplacian-iso}) to evaluate the derivatives in the CHE for the present simulations.
The simulation parameters are
$N_{L} = 40$, $Cn = 0.1$, $Pe = 8 \times 10^{3}$.
The temporal discretization parameter is
$N_{t} = 400$ ($c = 10$) for $Oh = 0.037$ and $0.119$,
and $N_{t} = 2000$ ($c = 50$) for $Oh=0.3$.

\begin{figure}[htp]
	\centering
	\includegraphics[trim= 1mm 1mm 1mm 1mm, clip, scale = 0.33, angle = 0]{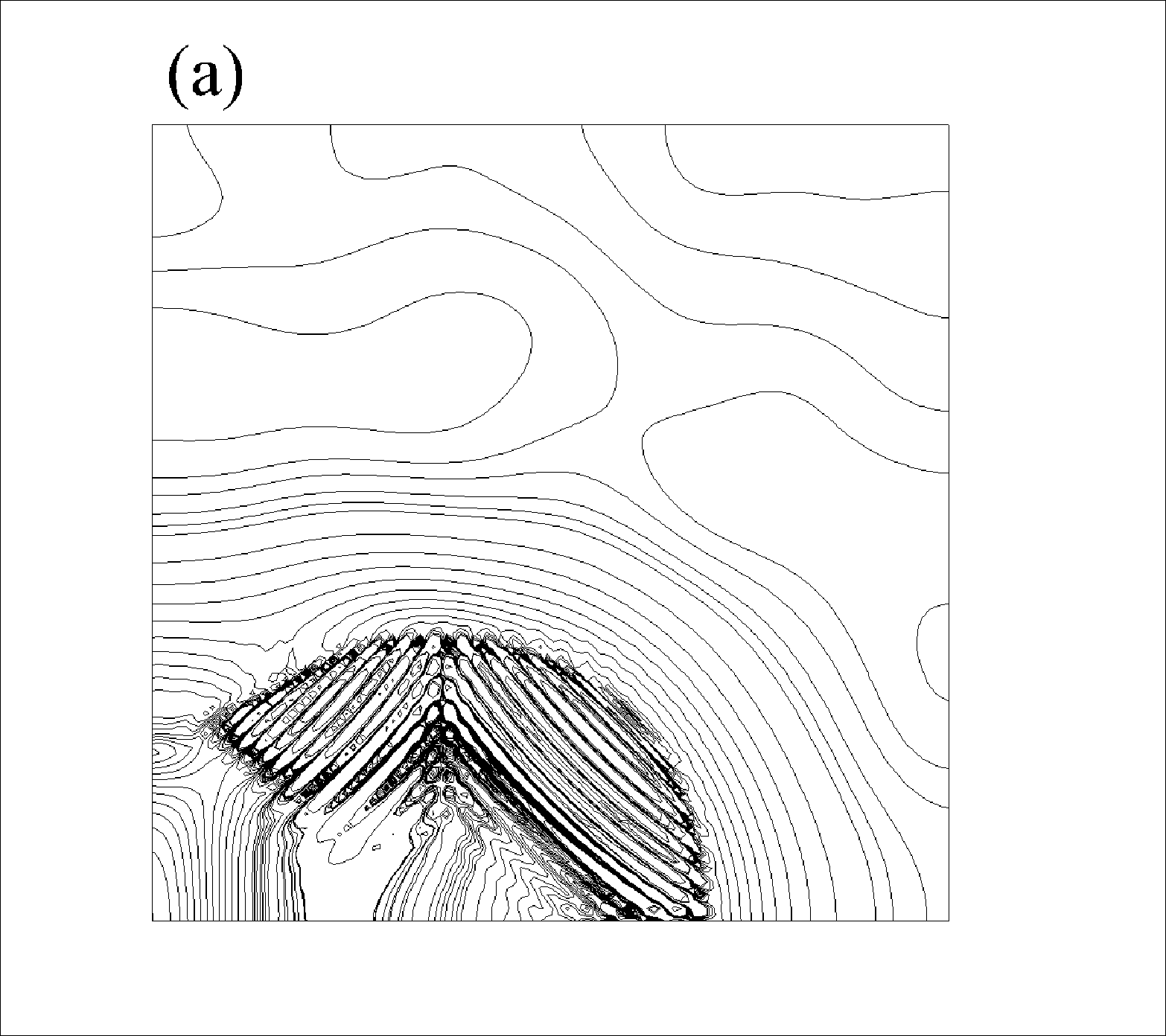}
	\includegraphics[trim= 1mm 1mm 1mm 1mm, clip, scale = 0.33, angle = 0]{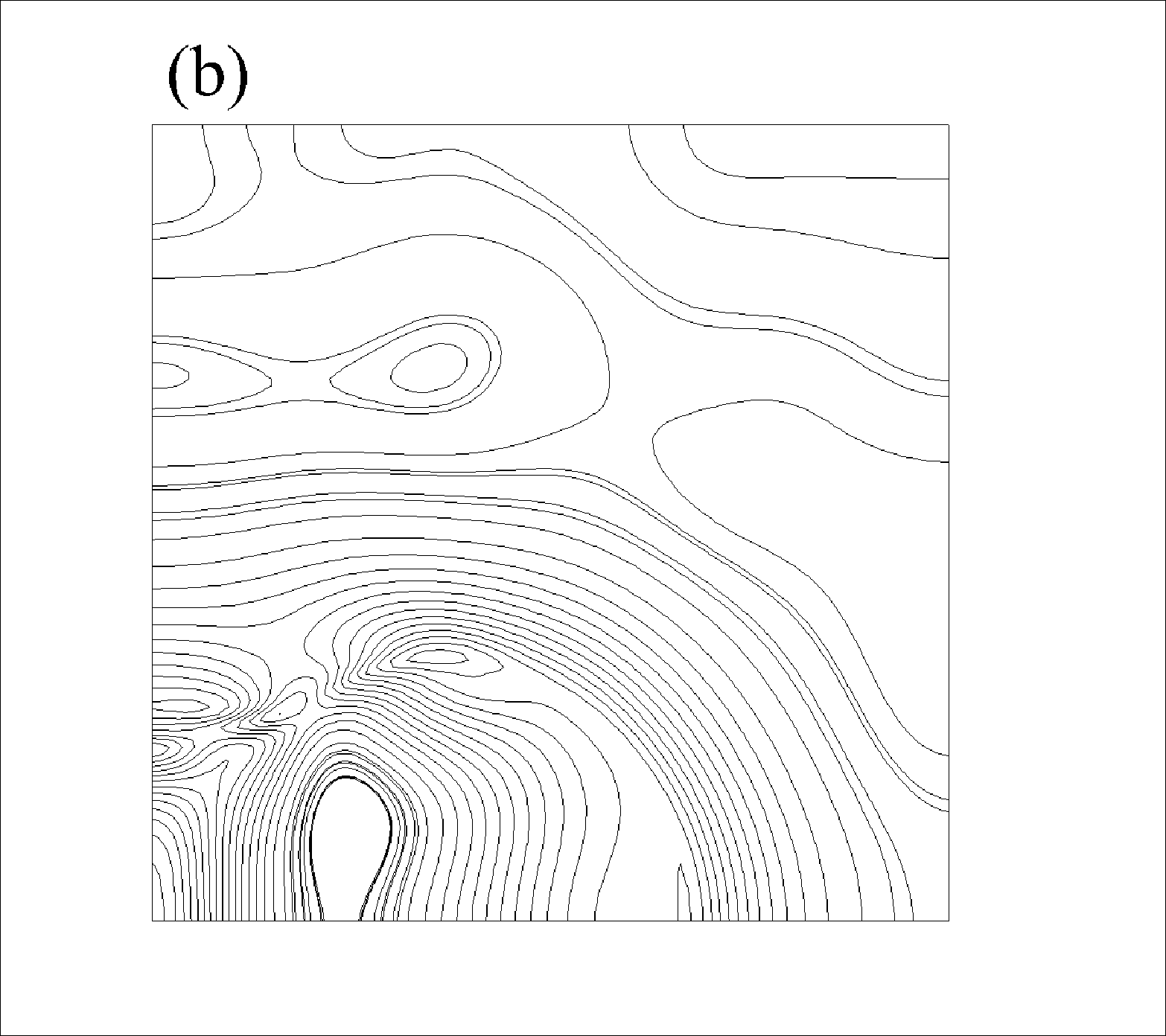}
	\caption{
		Pressure contours in the $x-z$ plane at $y=0$ at $t=0.4$ at $Oh= 0.037$ for the coalescence induced drop jumping on a nonwetting wall by the present simulation. In (a) the bulk viscosity is set to $\eta_{b} (\phi) = 0$ and in (b) it is set to be the same as the shear viscosity $\eta_{b} (\phi) = \eta (\phi) $.
	}
	\label{fig:drop-coal-3d-p-xz-oh0d037-t0d4}
\end{figure}

\begin{figure}[htp]
	\centering
	\includegraphics[trim= 1mm 1mm 1mm 1mm, clip, scale = 0.33, angle = 0]{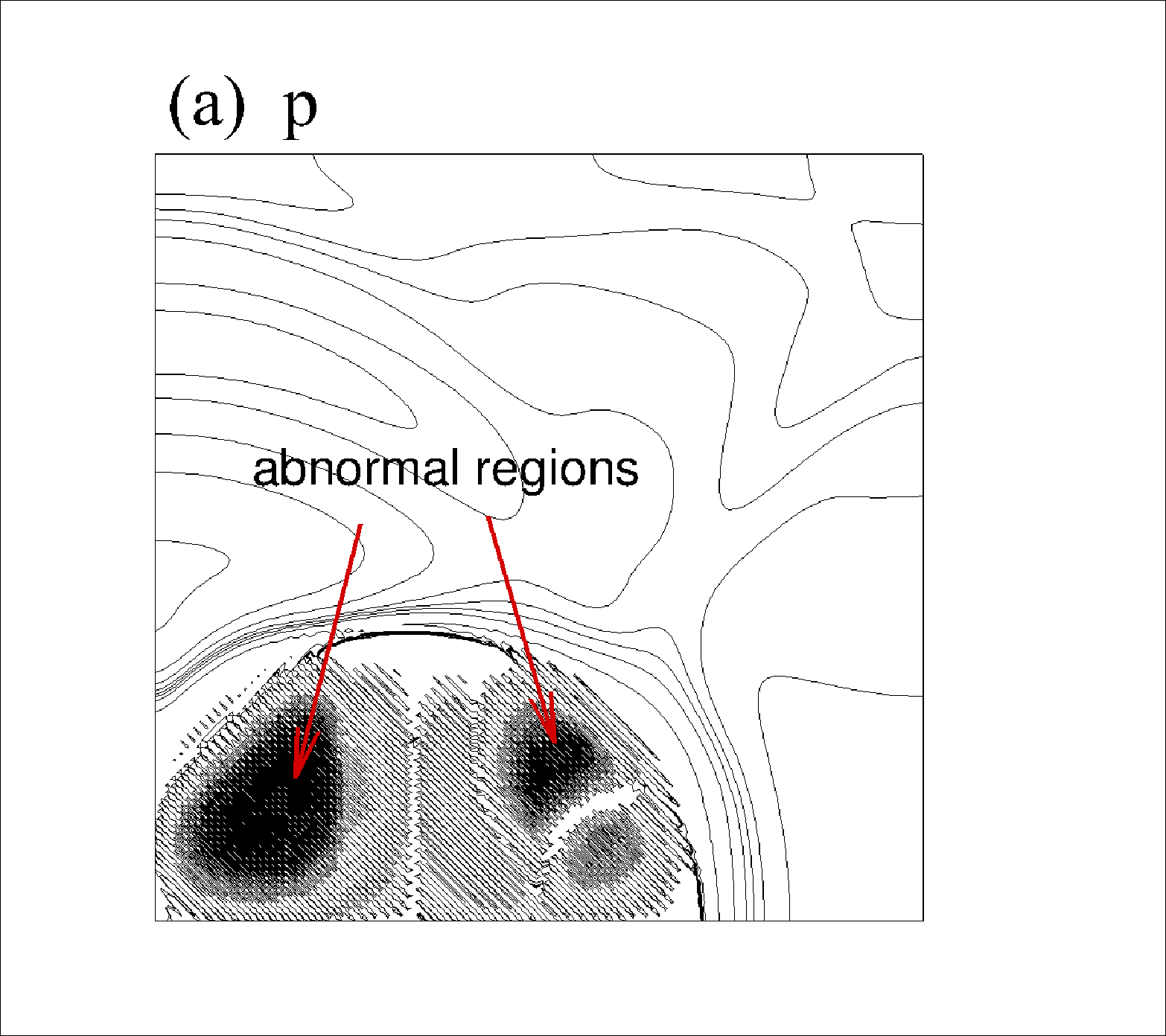}
	\includegraphics[trim= 1mm 1mm 1mm 1mm, clip, scale = 0.33, angle = 0]{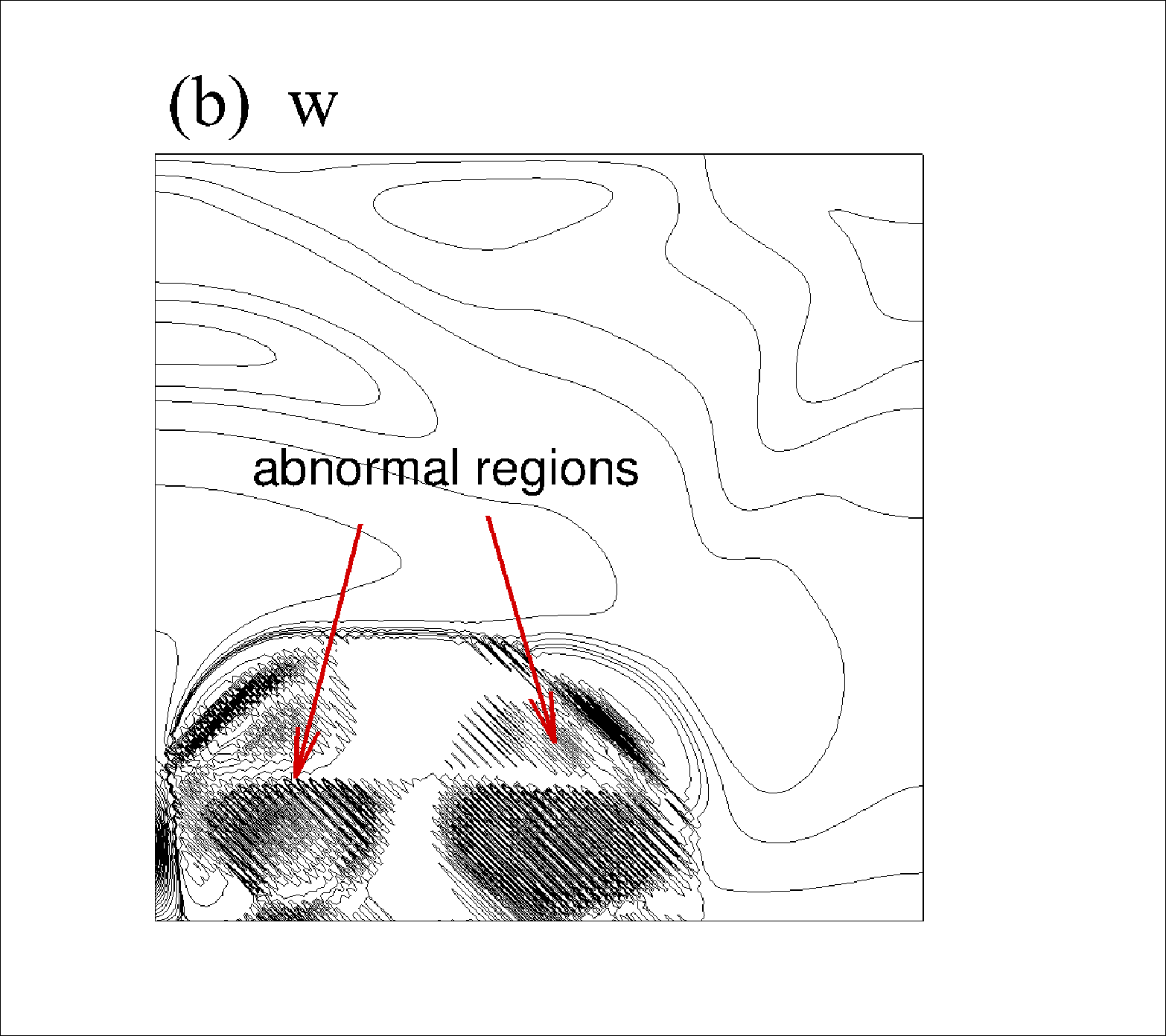}
	\caption{
		Contours of (a) the pressure $p$ and (b) the $z-$component of the velocity $w$ in the $x-z$ plane at $y=0$ at $t=0.021$ (when the simulation stopped due to instability) for the case at $Oh= 0.3$ for the drop coalescence in 3D by the present simulation. The simple second order scheme (without any averaging) was used to evaluate the pressure gradient terms.
	}
	\label{fig:drop-coal-3d-ff-xz-oh0d3-abnormal}
\end{figure}

Before presenting the results,
we briefly demonstrate the effects of the bulk viscosity (cf. eq. \ref{eq:vis-stress-two-phase})
and the way to calculate the pressure gradients (cf. eqs. \ref{eq:gradp-u-discretization} and \ref{eq:gradp-u-p-ave-3d}).
Based on our experience,
while these two factors have little effect in the 2D and axisymmetric simulations, 
they are crucial for 3D cases.
First, the effect of bulk viscosity was investigated for the case at  $Oh= 0.037$.
For this study, the pressure gradients were evaluated using suitable averages like eq. \ref{eq:gradp-u-p-ave-3d}.
Figure \ref{fig:drop-coal-3d-p-xz-oh0d037-t0d4} shows the pressure contours in the $x-z$ plane at $y=0$ at $t=0.4$ with the bulk viscosity $\eta_{b} (\phi) = 0$ and with $\eta_{b} (\phi) = \eta (\phi) $.
From fig. \ref{fig:drop-coal-3d-p-xz-oh0d037-t0d4}, one can see that with the bulk viscosity set to zero the pressure field shows checkerboard patterns at this time.
Besides, the flow fields have some wiggling variations in certain regions and the drop velocity $W_{drop}$ displays abrupt abnormal changes (not shown here).
By contrast, when the bulk viscosity is set to be the same as the shear viscosity, the checkerboard patterns disappear and the pressure field becomes relatively smooth. The drop velocity also evolves in a controlled manner as in the corresponding LBM simulation (to be shown below).
Second, the effect of pressure gradient evaluation was studied for the case at $Oh= 0.3$.
The bulk viscosity is set as $\eta_{b} (\phi) = \eta (\phi) $.
For this case, the simulation stopped due to instability problem at $t=0.021$
when the pressure gradients were calculated without any averaging (i.e., to set 
$\overline{p}_{i+\frac{1}{2},j} = p_{i,j,k}$ and $\overline{p}_{i-\frac{1}{2},j} = p_{i-1,j,k}$ in eq. \ref{eq:gradp-u-discretization}). Figure \ref{fig:drop-coal-3d-ff-xz-oh0d3-abnormal} shows the contours for the pressure and the $z-$component of the velocity $w$ in the $x-z$ plane at $y=0$ when the simulation stopped because of instability. We can also see checkerboard patterns here.
%and the pressure varies in an abnormally large range.
By contrast, when the averaged pressures (like eq. \ref{eq:gradp-u-p-ave-3d}) were used to evaluate the pressure gradients in the momentum equations, the simulation can finish without any problem and the results follow the corresponding LBM simulation closely (see below).

\begin{figure}[htp]
	\centering
		\includegraphics[trim= 1mm 1mm 1mm 1mm, clip, scale = 0.65, angle = 0]{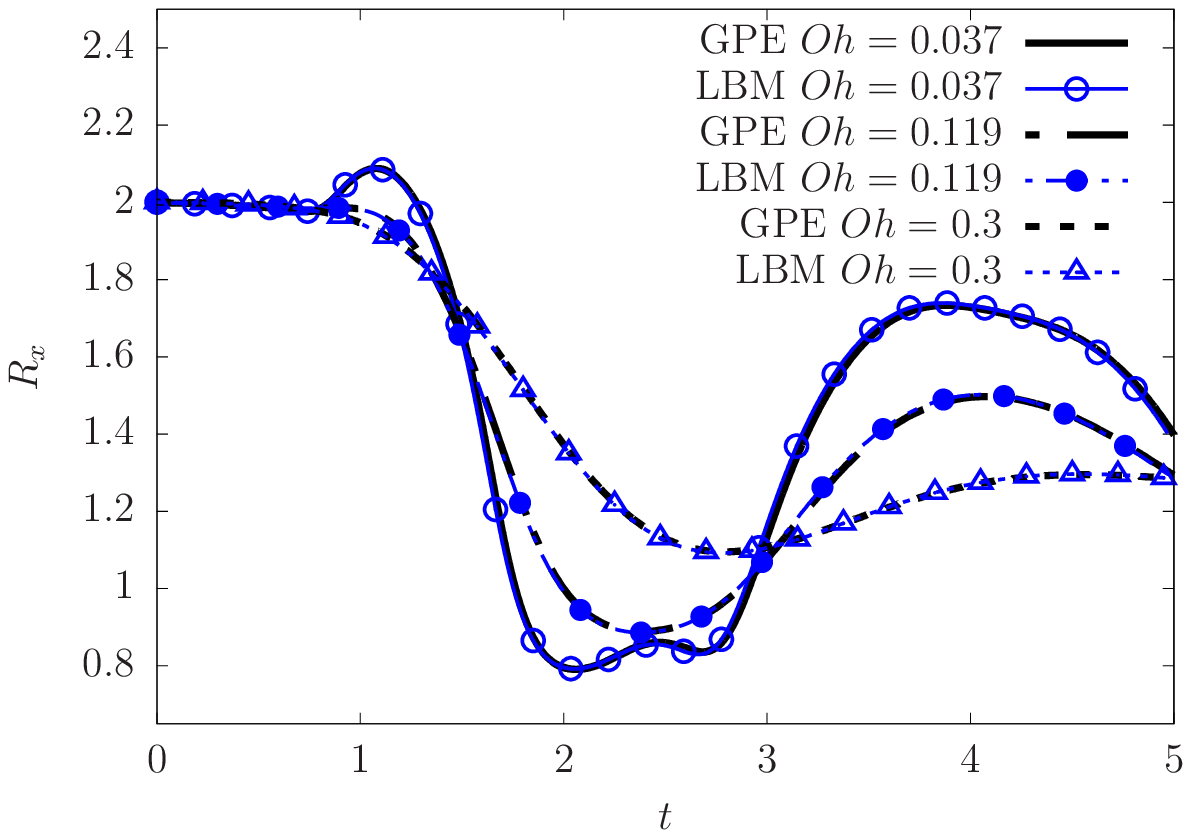}
		\includegraphics[trim= 1mm 1mm 1mm 1mm, clip, scale = 0.65, angle = 0]{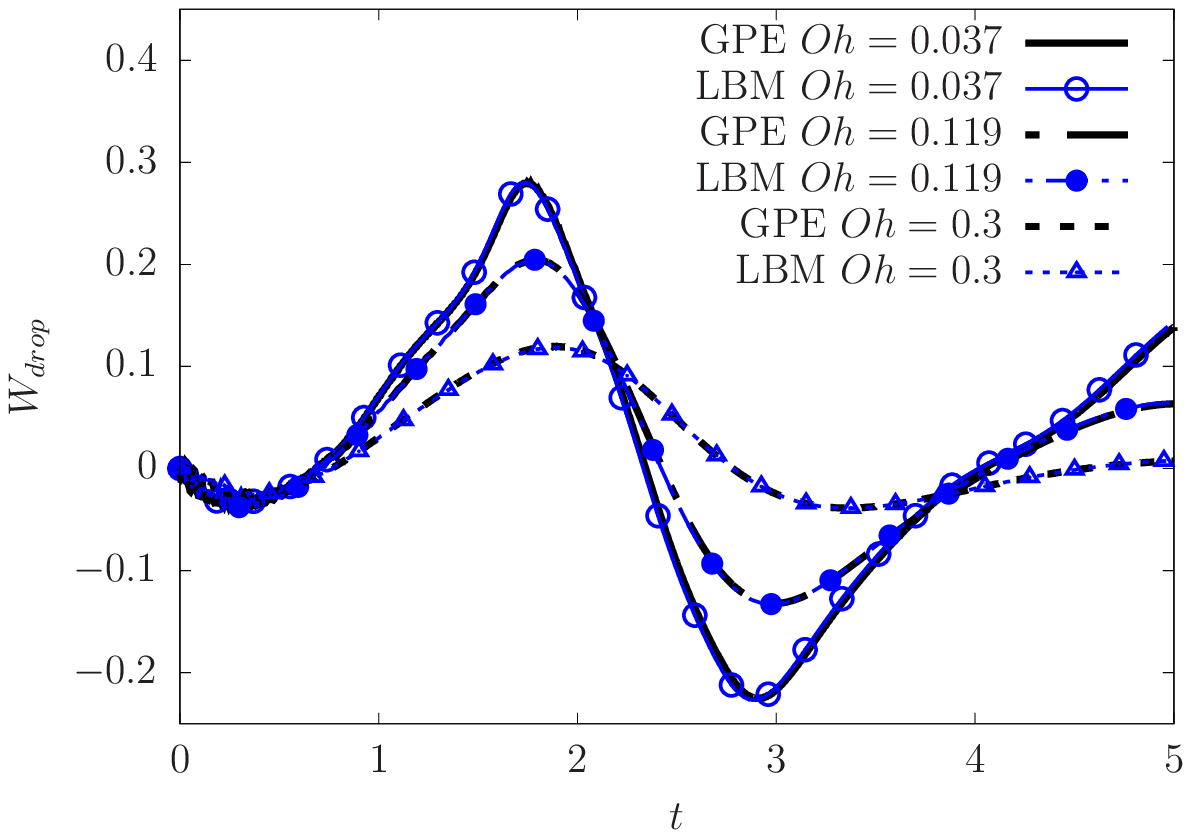}
	\caption{
		Evolutions of the (half) length of the drop in the $x-$direction at $y = z =0$ 
		and the $z-$component of the centroid velocity of the drop (in the simulation box, representing one eighth of the whole drop)
		during the coalescence of two drops by the present simulations.
		The shared simulation parameters are
		$N_{L} = 40$, 
		$Cn = 0.1$, $Pe = 8 \times 10^{3}$.
		The temporal discretization parameter is
		$N_{t} = 2000$ ($c = 50$) for $Oh=0.3$, and	$N_{t} = 400$ ($c = 10$) for $Oh=0.119$ and $0.037$.
	}
	\label{fig:drop-coal-3d-rx-udrop-evol}
\end{figure}

\begin{figure}[htp]
	\centering
	\includegraphics[trim= 5mm 5mm 5mm 5mm, clip, scale = 0.65, angle = 0]{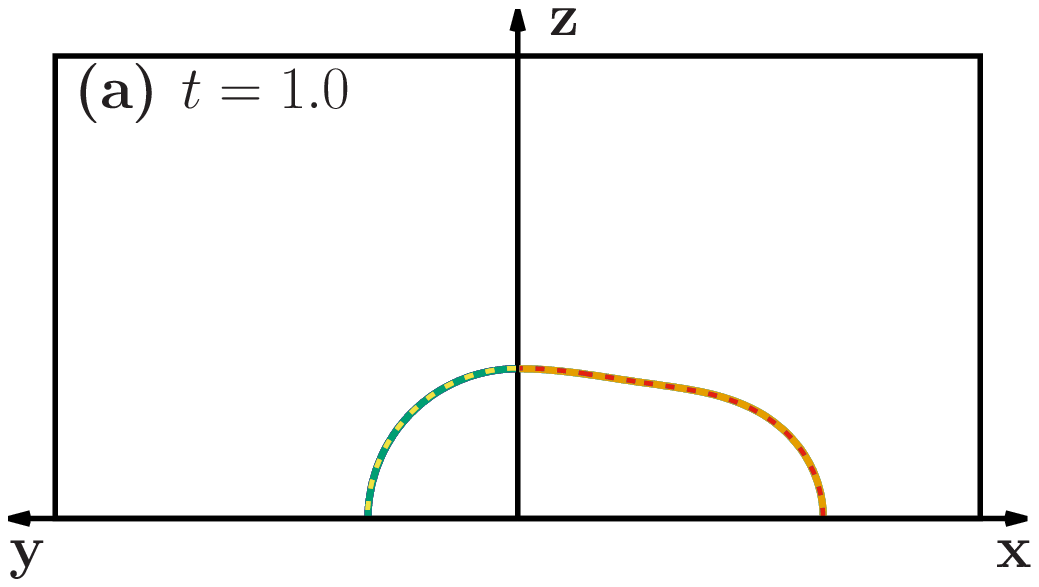}
	\includegraphics[trim= 5mm 5mm 5mm 5mm, clip, scale = 0.65, angle = 0]{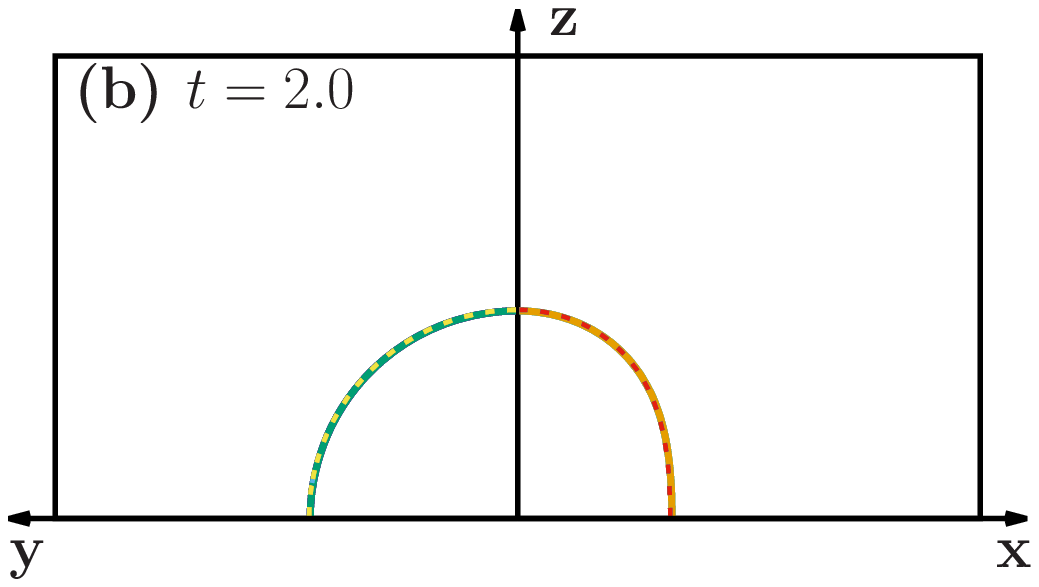}
	\includegraphics[trim= 5mm 5mm 5mm 5mm, clip, scale = 0.65, angle = 0]{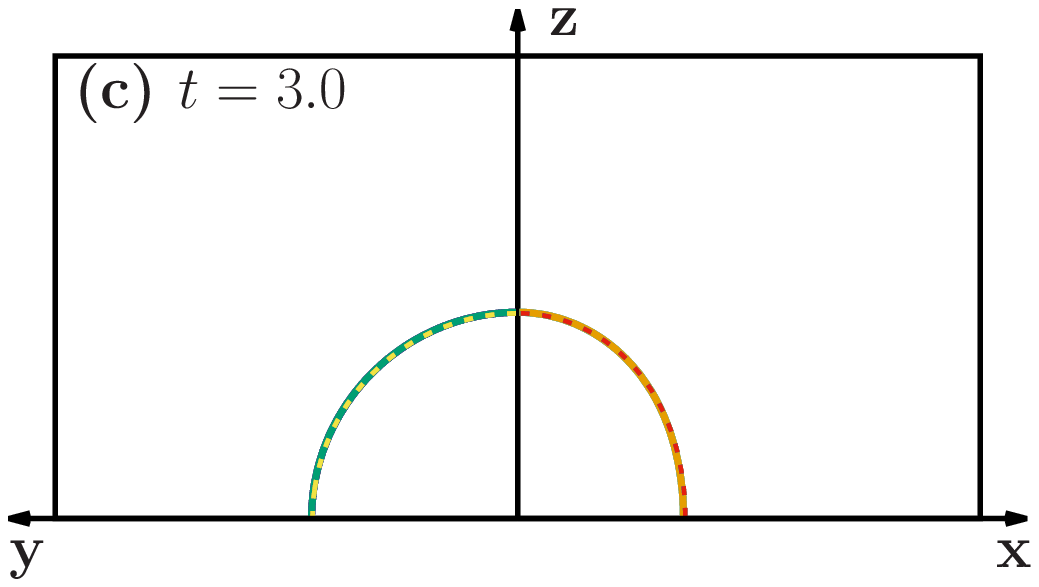}
	\includegraphics[trim= 5mm 5mm 5mm 5mm, clip, scale = 0.65, angle = 0]{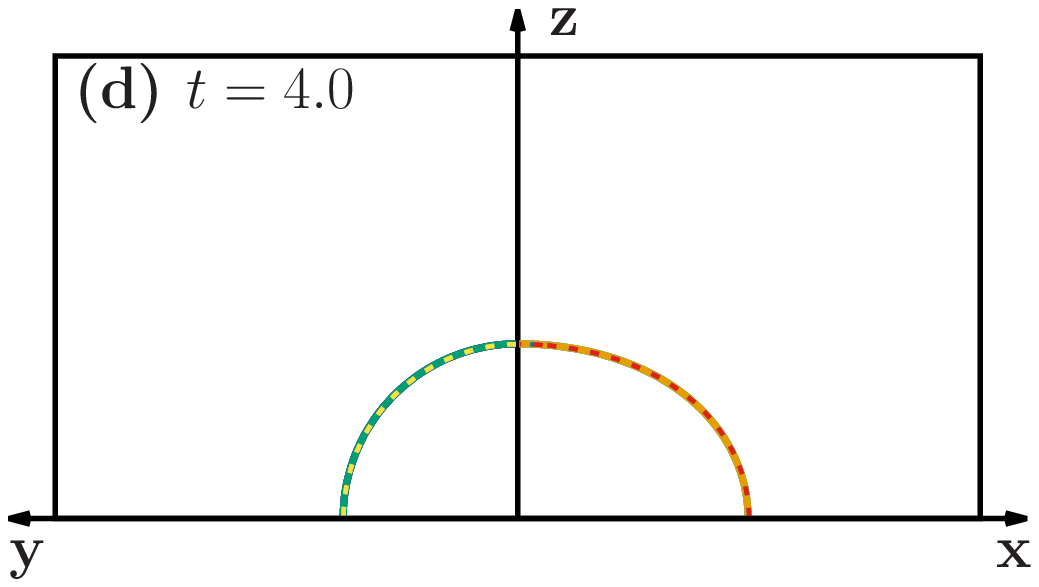}
	\caption{
		Snapshots of the interfaces at $t= 1.0$, $2.0$, $3.0$ and $4.0$ for the 3D drop coalescence 
		at $Oh= 0.119$
		by the present GPE-based simulation (solid) and by the LBM simulation (dashed).
		In each panel, the left half shows the $y-z$ plane at $x=0$ and the right shows the $x-z$ plane at $y = 0$.
	}
	\label{fig:drop-coal-3d-itf-t01-02-03-04}
\end{figure}

\begin{figure}[htp]
	\centering
	\includegraphics[trim= 1mm 1mm 1mm 1mm, clip, scale = 0.33, angle = 0]{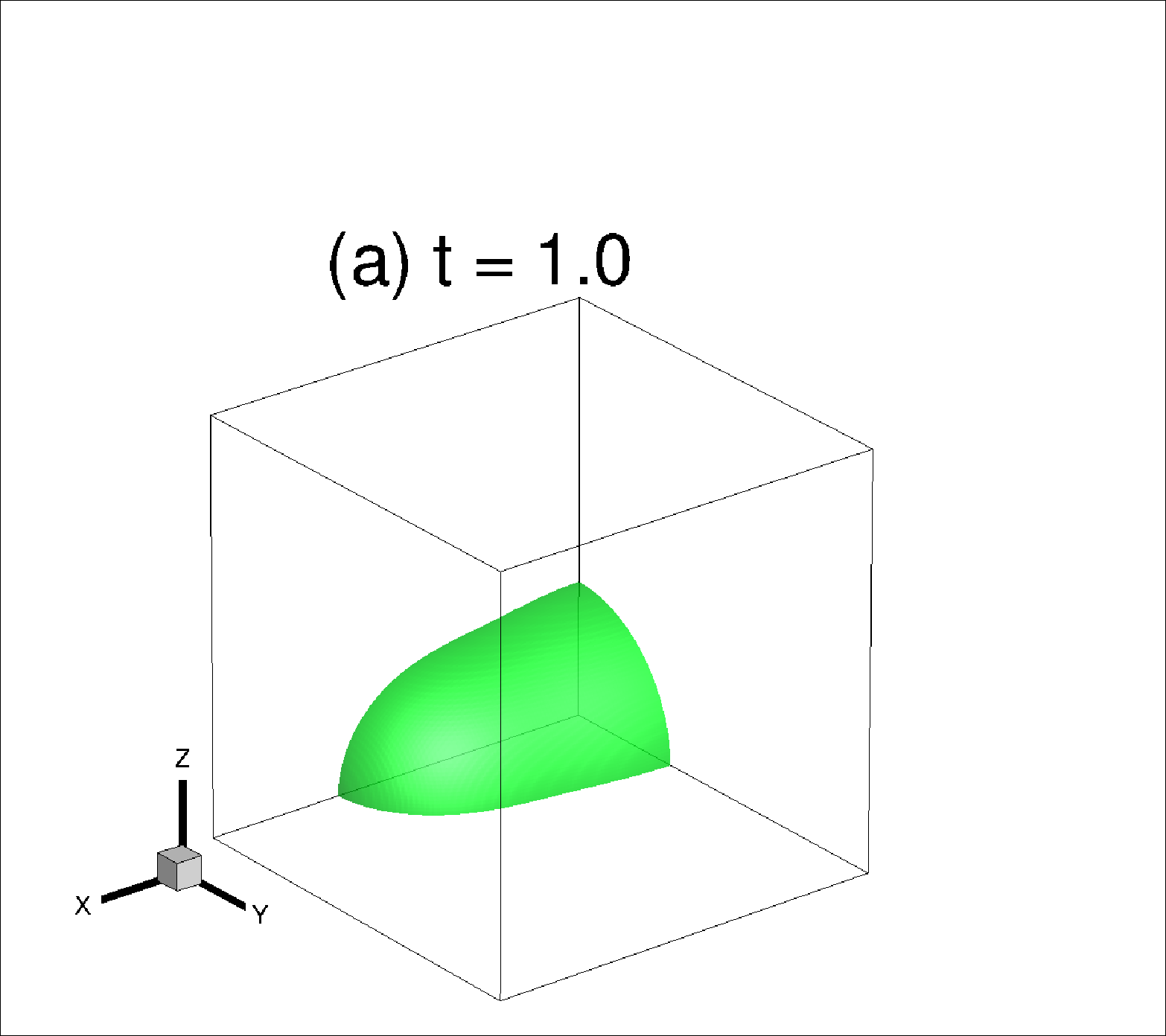}
	\includegraphics[trim= 1mm 1mm 1mm 1mm, clip, scale = 0.33, angle = 0]{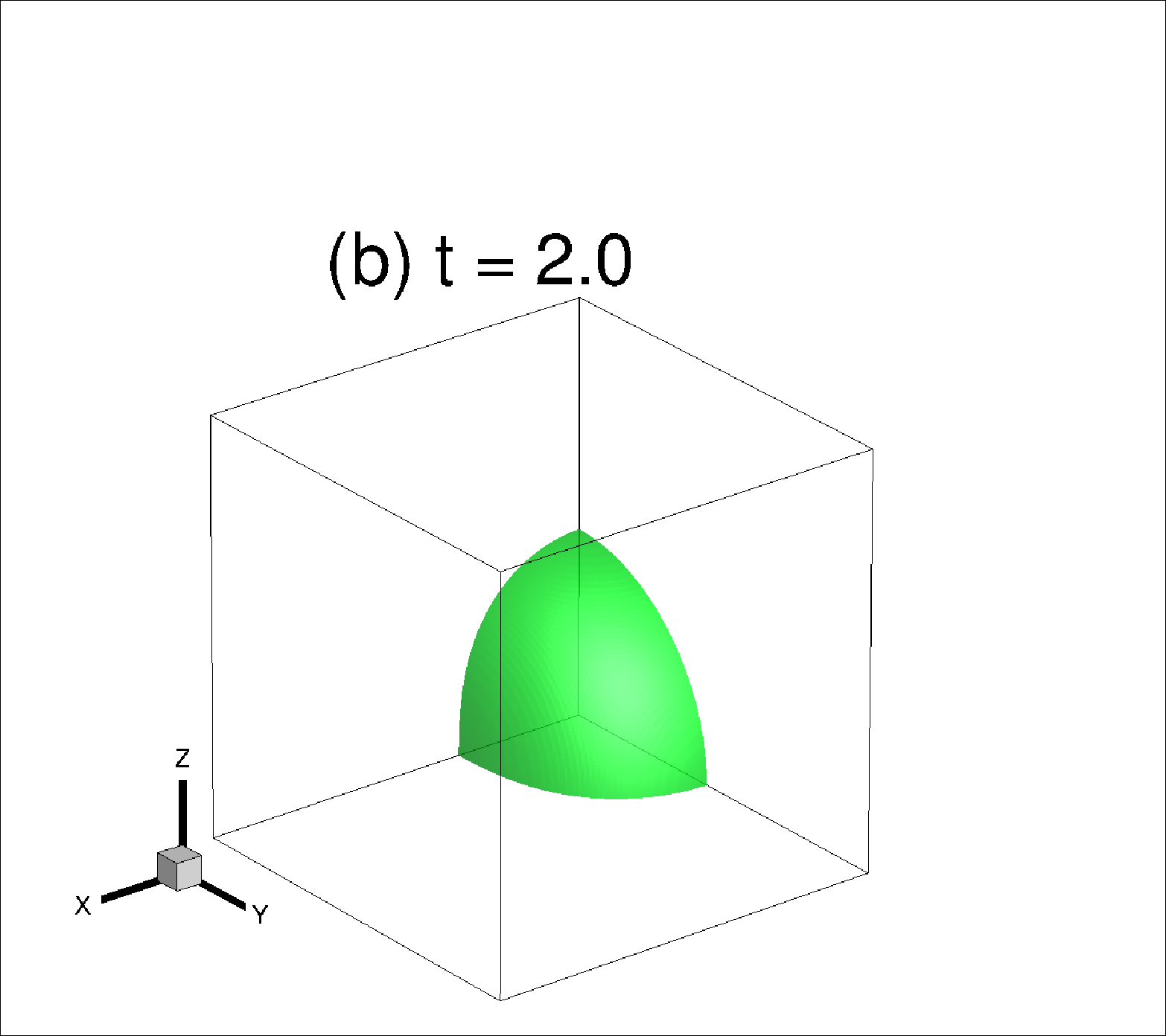}
	\includegraphics[trim= 1mm 1mm 1mm 1mm, clip, scale = 0.33, angle = 0]{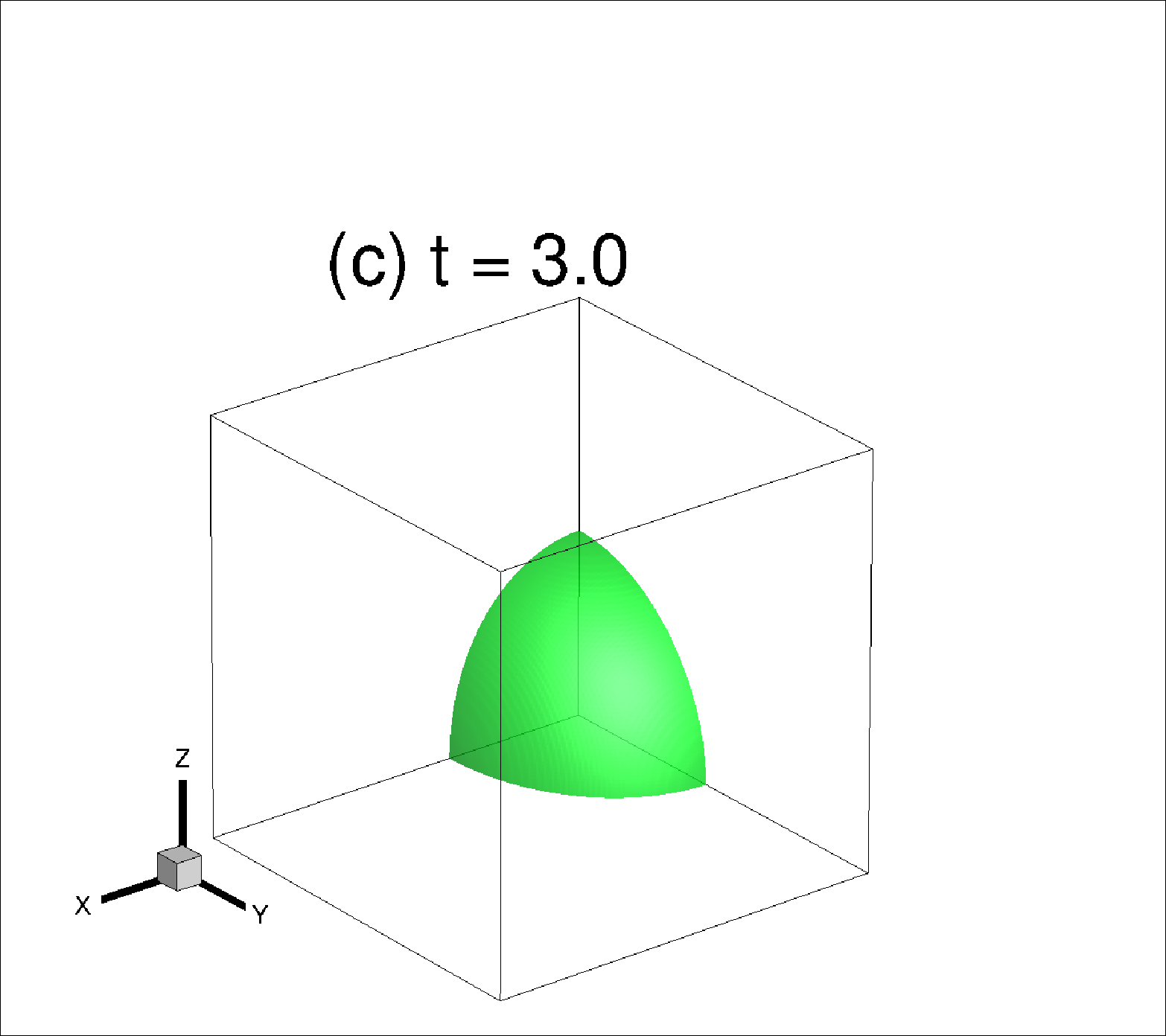}
	\includegraphics[trim= 1mm 1mm 1mm 1mm, clip, scale = 0.33, angle = 0]{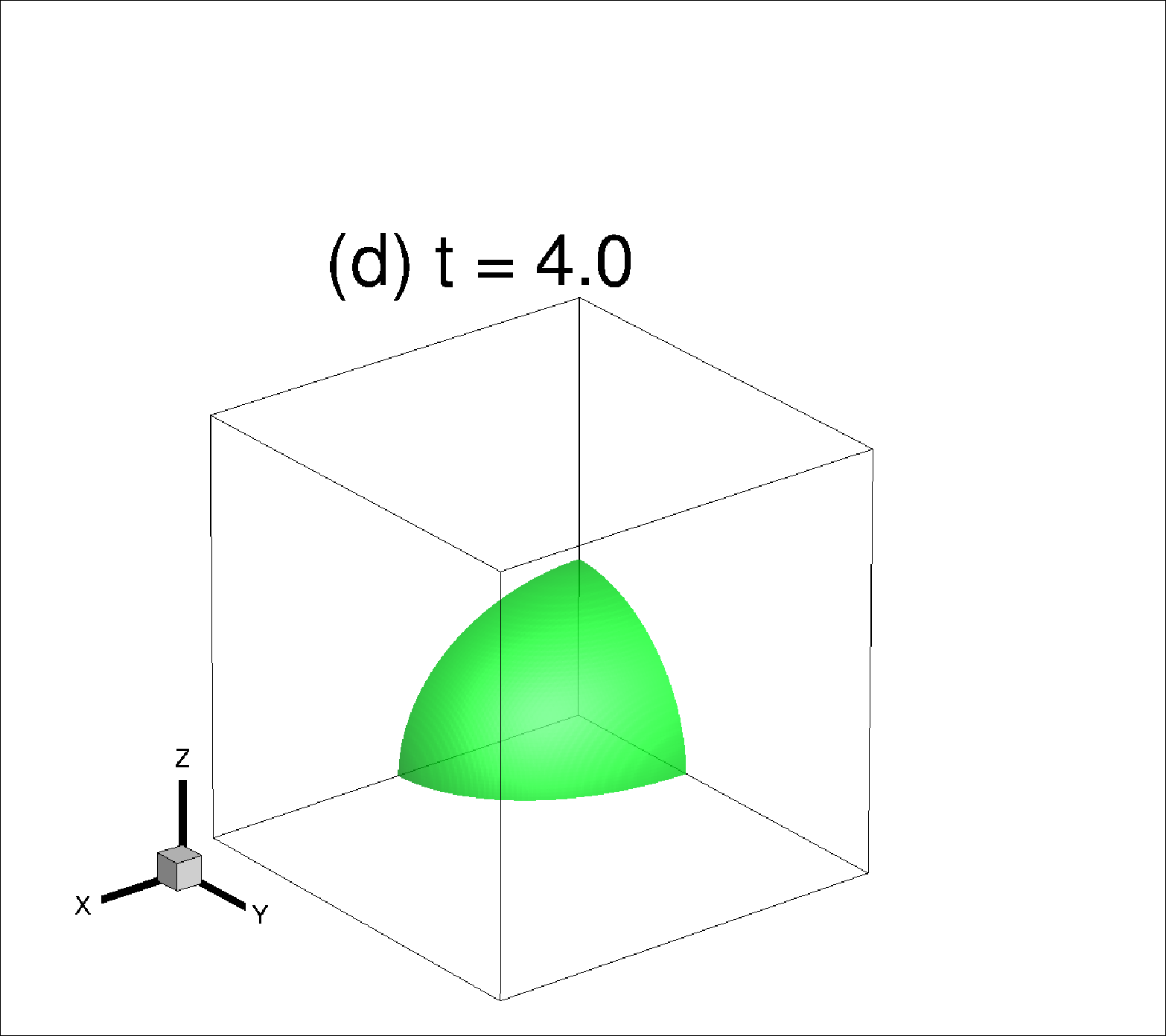}
	\caption{
		Snapshots of the drop in 3D at $t= 1.0$, $2.0$, $3.0$ and $4.0$ at $Oh= 0.119$ for the drop coalescence by the present simulation.
	}
	\label{fig:drop-coal-3d-t01-02-03-04}
\end{figure}

Now the results obtained by the normal simulations with $\eta_{b} (\phi) = \eta (\phi) $ and pressure averaging for the gradient calculation are presented.
Figure \ref{fig:drop-coal-3d-rx-udrop-evol} shows the evolutions of the (half) length of the drop along the $x-$axis $R_{x}$ and
the centroid velocity of the drop (one eighth of the whole drop) in the $z-$direction $W_{drop}$ for the three cases at $Oh = 0.037$, $0.119$ and $0.3$.
Note that the centroid of the whole drop does not move due to symmetry. 
However, the part in the simulation box (one eighth of the whole domain) 
may have a nonzero velocity $W_{drop}$ during the coalescence process.
It can be found that the present results are in good agreement with 
those obtained by the LBM for all cases.
Besides, the interfaces in the symmetry planes (the $y-z$ plane at $x=0$ and the $x-z$ plane at $y = 0$) were examined and it was found that the present results are quite close to the reference ones as well. 
For brevity, we only show the interfaces in the symmetry planes
at four selected times for the case at $Oh=0.119$ 
in fig. \ref{fig:drop-coal-3d-itf-t01-02-03-04}.
Several 3D snapshots of the drop for this case 
are given in fig. \ref{fig:drop-coal-3d-t01-02-03-04} to better illustrate the evolution of drop morphology.

\subsection{Comparisons of computation time and memory requirement}\label{ssec:cmp-gpe-lbm}

Finally, we compare the computation time and memory requirement by the GPE-based and LBM simulations.
One case in Section \ref{ssec:drop-coal-3d}
at $Oh = 0.119$ is selected for this comparison.
The discretization parameters are the same ($N_{L} = 40$, $N_{t} = 400$).
The grid is $120 \times 120 \times 120$ and the total number of steps are $8$. 
Both codes were run in parallel using four computational nodes (with the domain decomposed into four boxes of the same size in the $x-$direction).
Three different velocity models were tested: D3Q15, D3Q19 and D3Q27.
As the spatial derivatives in the CHE were evaluated using the isotropic schemes which depend on the specific velocity model, the choice of the velocity model also affects the GPE-based simulation results to some extent.
But the memory requirement is not affected because no distribution functions are used in such simulations.
The simulation time is given in Table \ref{tab:cmp-gpe-lbm-time}.
It can be found that the GPE simulations are slower than the LBM simulations, but the difference becomes smaller as the number of lattice velocities increases.
When D3Q15 is used, the LBM simulation costs about half of the time by the GPE simulation.
When D3Q27 is used, the GPE simulation roughly takes $50 \%$ more time than the LBM one.
Therefore, when one needs to obtain the results in a shorter time, the LBM is recommended.
At the same time, the GPE simulation only requires to store $p$, $u$, $v$, $w$, $\phi$ and $\mu$.
In contrast, the LBM simulation has to store at least additional $b$ 
($b=15$, $19$ and $27$ for D3Q15, D3Q19 and D3Q27) distribution functions.
Thus, the memory requirement by the LBM is at least $3.5$, $4.2$ and $5.5$ times higher than that by the GPE for D3Q15, D3Q19 and D3Q27 respectively.
If one has to deal with a large scale problem but with only limited memory,
the GPE-based method is preferred.
Lastly, it is noted that the GPE-based simulation uses the TVD-RK3 scheme (eq. \ref{eq:tvd-rk3}) and to march one time step it is necessary to perform the variable updates for three times.
If only one update per step was enough, the GPE-based method would be faster than the LBM.
But in our tests, it was found that the TVD-RK3 scheme is necessary to keep the simulation stable (even the TVD-RK2 scheme did not suffice).

\begin{table}[htp]
	\caption{Computation time used by the GPE-based and LBM simulations for the same case using same numerical parameters.}
	\label{tab:cmp-gpe-lbm-time} 
	\begin{center}
		\begin{tabular}{|c|c|c|c|}\hline
			Velocity model & D3Q15 & D3Q19 & D3Q27\\\hline
			Time by GPE (s) & 19.1 & 19.5 & 19.7 \\\hline
			Time by LBM (s) & 10.3 & 11.3 & 13.5 \\\hline
			Ratio (GPE/LBM)& 1.9 & 1.7 & 1.5 \\\hline
		\end{tabular}
	\end{center}
\end{table}

\section{Concluding Remarks}\label{sec:conclusion}

To summarize, 
we have proposed a numerical method using the general pressure equation for the simulation of two-phase incompressible viscous flows with variable density and viscosity. 
It resembles the lattice Boltzmann method in that the incompressibility condition is relaxed to some extent to allow certain compressibility.
It was verified through a number of tests including the capillary wave and Rayleigh-Taylor instability in 2D, 
the falling drop and drop coalescence under axisymmetric geometry,
and the drop coalescence in 3D.
All the results are in good agreement with reference results either by the LBM or by other methods solving the Poisson equation in the literature.
The most attracting feature of the present method is that it uses much less memory than the LBM, especially in 3D.
As of now, it is still difficult for the present method to deal with two-phase flows with large density ratios in its current formulation.
In future, more sophisticated discretization schemes (e.g., upwind schemes for the convection terms)
may be used  to enhance its capability to handle more challenging problems.
Besides, the phase-field model may be refined by including some other development (e.g.,~\cite{jcp11-conserv-pf, pof19-itf-compressed}).

\bf Acknowledgement \rm

This work is supported by the National Natural Science Foundation of China 
(NSFC, Grant No. 11972098).

\bf Data availability \rm

The data that support the findings of this study are available from the corresponding author upon reasonable request.

\bf Declaration of interests \rm

The authors declare that they have no known competing financial interests or personal relationships that could have appeared to influence the work reported in this paper.

%\bibliography{/Users/jjhuang/work/bib/MyReference}

\end{document}